\documentclass[%
reprint,
 amsmath,amssymb,
 aps,
prl,
]{revtex4-2}

\usepackage{xcolor}
\usepackage{graphicx}
\usepackage{dcolumn}
\usepackage{physics}
\usepackage{subfig}
\usepackage{xr}
\usepackage{hyperref}
\usepackage{bm}

\begin{document}

\preprint{APS/123-QED}

\title{Analysis  of  the blowout plasma wakefields produced by drive beams with elliptical symmetry}

\author{P. Manwani}
\thanks{pkmanwani@gmail.com}
\author{Y. Kang}
\author{J. Mann}
\author{B. Naranjo}
\author{G. Andonian}
\author{J. B. Rosenzweig}
\affiliation{%
Department of Physics and Astronomy, UCLA, Los Angeles, California 90095, USA
}%

\date{\today}

\begin{abstract}
In the underdense or blowout regime of plasma wakefield acceleration, the particle beam is denser than the plasma.  In this scenario, the plasma electrons are nearly completely rarefied from the beam channel, leaving only a nominally  uniform ion-filled "bubble". Extensive investigations of this interaction assuming axi-symmetry have been undertaken. However, the blowout produced by a transversely asymmetric (flat) driver, which would be present in linear collider "afterburner" schemes, possesses quite different characteristics. Such beams create an asymmetric plasma bubble which leads to unequal focusing in the two transverse dimensions, accompanied by a non-uniform accelerating gradient. The asymmetric blowout cross-section is found through simulation to be elliptical, and treating it as such permits a simple extension of the symmetric theory. In particular, focusing fields linear in both transverse directions inside the bubble are found. The form of the wake potential and the associated beam matching conditions in this elliptical cavity are discussed. We also examine blowout boundary estimation in the long driver limit and applications of the salient asymmetric features of the wakefield.
\end{abstract}

\maketitle



Plasma wakefield acceleration (PWFA) is an  emerging technique for high-energy accelerator applications, including future particle colliders \cite{Lindstrom_2021} \cite{Positrons2024} and free-electron lasers \cite{Pompili_2022}. By using electromagnetic fields many orders of magnitude stronger than those in current linear accelerators, PWFA may enable compact acceleration schemes reaching TeV-scale energies in a few 100's of meters.  The far-reaching application of this approach is to create high-luminosity $e^+e^-$  colliders.  To minimizes strong beam-beam forces and radiative energy spread (beamstrahlung) \cite{chen} at the interaction point \cite{ilc_2013,clic_2012}, these colliders use highly asymmetric beams with associated large transverse emittance ratio $\epsilon_{n,x}/\epsilon_{n,x}$Further, the most viable path to the first demonstration of using PWFA to push the energy frontier involves integrating plasma acceleration as a ``afterburner" \cite{energy_doubler}. This would entail using high  $\epsilon_{n,x}/\epsilon_{n,x}$ beams to drive the wakefields. In this scenario, both the accelerating and the driving beams would have inherent $x$-$y$  asymmetries -- the beam will tend to be transversely flat.  Although many PWFA-based collider scenarios have been proposed, despite the essential relevance of the afterburner concept  the study of the flat beams implied in the PWFA in its highly likely first incarnation has not yet been notably developed. Similar comments can be made concerning other, related schemes, such as powerful plasma-based asymmetric final-focus schemes \cite{tor_afterburner,Weidemann_2002}, which may importantly include an adiabatic plasma focusing section\cite{AdiabaticPlasmaLens} to avoid the limitations of radiation-induced aberrations on the interaction quality\cite{Oide_1988}. Thus, the plasma response and associated dynamics and evolution of flat beam properties, which are the central themes of this Letter, are ripe for investigation. This study, due to its extension of relevant has impacts on fields ranging from fundamental plasma and beam physics, to nonlinear dynamics, and on to energy-frontier particle physics. 

While the PWFA was initially proposed in the linear regime \cite{PChen1985}, where it is based on plasma electron oscillations harmonic at the plasma frequency $\omega_p=\sqrt{e^2n_0/\epsilon_0 m_e}$ ($n_0$ is the nominal plasma density), it is now most commonly used in the nonlinear, or blowout, regime \cite{jamie2d_1991,rosenzweig_2000}. In the blowout case, where the beam density $n_b\gg n_0$, the strong electric fields of the driver expel the plasma electrons outward, creating a blowout bubble devoid of electrons. The expelled electrons and the electrons within a plasma skin-depth of the boundary form a dense electron sheath which envelops the cavity. This highly localized plasma electron density region and the associated return current both serve to shield the drive beam's electromagnetic (EM) fields outside of the bubble \cite{jamie2d_1991, weilu_2006, yi_2013}. The forces acting on the beam inside this axisymmetric bubble are quite ideal for acceleration, with no dependence of the longitudinal wake forces on transverse offset $r$. Further, due to the uniform ion column inside the bubble, the beam electrons undergo geometric-aberration-free focusing dependent linearly on  $r$. While the axisymmetric PWFA has been extensively studied (see, \textit{e.g.} \cite{Qtilde1} and \cite{Qtilde2}), there are numerous open questions concerning the physics of wake waves driven by asymmetric drivers \cite{Baturin_2022}. Indeed, with the breaking of axisymmetry, now and potentially deleterious effects on beam propagation \cite{deiderichs_2024} can develop.  With these motivations, here we take significant steps forward in addressing the theory of asymmetric beams in the PWFA.  

To proceed with the analysis of flat-beam scenarios, we use the three-dimensional (3D) particle-in-cell (PIC) code OSIRIS \cite{Fonseca_2008}. Simulations indicate that the blowout created by flat beams can be well-approximated as elliptical in cross-section.  Consequently, the potential inside these elliptical cavities, which translate in $z$ at nearly the speed of light ( $v_b\simeq c$), depends quadratically on $x$ and $y$, yielding linear transverse electric fields differing in strength along the two transverse axes \cite{manwani_ipac_22,manwani_aac}. This plasma lens focusing asymmetry provides unique properties which may yield a key advantage in a final focusing system for next-generation colliders \cite{thin_plasma_lens_doss_2019,tor_afterburner,rajagopalan_1995}, enhancing the desired asymmetry at the interaction point. 

In this Letter we use normalized units, where densities are normalized to $n_0$, which then specifies $\omega_p$.  With this fundamental time-scale given, a unitless notation is implemented:  time is normalized to $\omega_p^{-1}$; velocities to the speed of light $c$; distances to the plasma skin-depth $k_p^{-1}=c/\omega_p$; velocity to $c$; charge to $e$; EM field amplitudes to the so-called wave-breaking \cite{Dawson_1959} value,  $E_{wb} = m_e c\omega_p/e$; and potentials to $m_e c^2/e$. Source terms and their species  are indicated using the following subscripts: the plasma ions (i), the plasma electrons (\textit{e}), and the drive beam (\textit{b}). To simplify our analysis, we assume that the massive ions are static, forming a uniform background charge of $n_i=n_0=1$. This assumption holds when the oscillation phase advance of the ions in the potential of the beam is small, $\Delta \phi=\sigma_z \sqrt{\pi Z_i n_b/m_i} \ll 1$, where $Z_i$ is the ionization state of the ions, $m_i$ is the ion mass, and $\sigma_z$ is the beam bunch length   \cite{ionmotion_rosenzweig_2005}. The EM source terms are then given by: the charge density $\rho = \rho_b + \rho_e + 1$ and the current density $\bm{J} = \bm{J_{b}} + \bm{J_{e}}$. 

The equations of motion for the plasma electrons can be written in Hamiltonian form by using the vector and scalar potentials $\bm{A}$ and $\phi$, and the canonical momenta $\bm{P}=\bm{p}+\bm{A}$. The beam evolution occurs on a much longer time-scale than  the wakefield in the $\xi$-frame, allowing use of the quasi-static (slowly-varying in $s$, $\partial_s \ll \partial_\xi$) approximation, $(x, y, z, t) \rightarrow (x, y, \xi \equiv t - z, s \equiv z)$ \cite{sprangle_1990},  
.

With these assumption Maxwell’s equations for the normalized potentials in the Lorentz gauge reduce to
\begin{equation}
 \nabla_\perp^2 \left[\begin{array}{c}
\phi \\
\bm{A}
\end{array}\right] = -\left[\begin{array}{c}
\rho \\
\bm{J}
\end{array}\right] ,
\label{scalar_vector}
\end{equation}
where $\nabla_\perp^2=\partial_x^2 + \partial_y^2$ is the transverse Laplace operator. The Lorentz gauge condition $\bm{\nabla} \cdot {\bm{A}}+\frac{\partial \phi}{\partial t}=0$ can then be written as $\bm{\nabla}_{\perp} \cdot \bm{A}_{\perp}=-\frac{\partial}{\partial \xi}\left(\phi-A_z\right)=-\frac{\partial}{\partial \xi} \psi$.  Here $\psi=\phi-A_z$ is the wake potential, which obeys the transverse Poisson equation $-\nabla_{\perp}^2 \psi=\rho-J_z=S$. Here we have defined a source density $S$. The continuity equation in the $\xi$-frame can be written as $\frac{\partial}{\partial \xi}S + \bm{\nabla}_\perp \cdot \bm{J_\perp} = 0$. Now, the Hamiltonian of an electron is given by $H={\gamma} -\phi$ with the Lorentz factor ${\gamma}=\sqrt{1+|\bm{p}|^2}$. In the quasi-static approximation,  and $H$ depends on $z$ and $t$ only through $\xi$. The time derivative of the Hamilton, $\frac{d H}{d t}=\frac{\partial H}{\partial t}=\frac{\partial H}{\partial \xi}=- \frac{\partial H}{\partial z}= \frac{d P_z}{d t}$, yields conservation of $H- P_z$. If the plasma  electrons are initially at rest, this implies, $\gamma-\phi- p_z+ A_z=1$  \cite{mora_1997,weilu_2006_pop}. Utilizing  the wake potential, this initial condition gives
\begin{equation}
    \gamma-\psi-p_z=1 .
    \label{eq:energy}
\end{equation}
The fields are found from the potentials as follows:
\begin{equation}
\begin{split}
    E_z &= \pdv{\psi}{\xi} ; \ \bm{E}_\perp = -\bm{\nabla}_\perp \phi - \pdv{\bm{A}_\perp}{\xi}; \\
    \bm{B_\perp} &= \bm{\nabla}_\perp \times \bm{A_z} +  
 \bm{\nabla}_z \times \bm{A_\perp} ; \ 
    B_z = \bm{\bm{\nabla}_\perp} \times \bm{A}_\perp
\end{split}
\label{eq:maxwell}
\end{equation}

\begin{figure}
    \centering
    \includegraphics[width=\columnwidth]{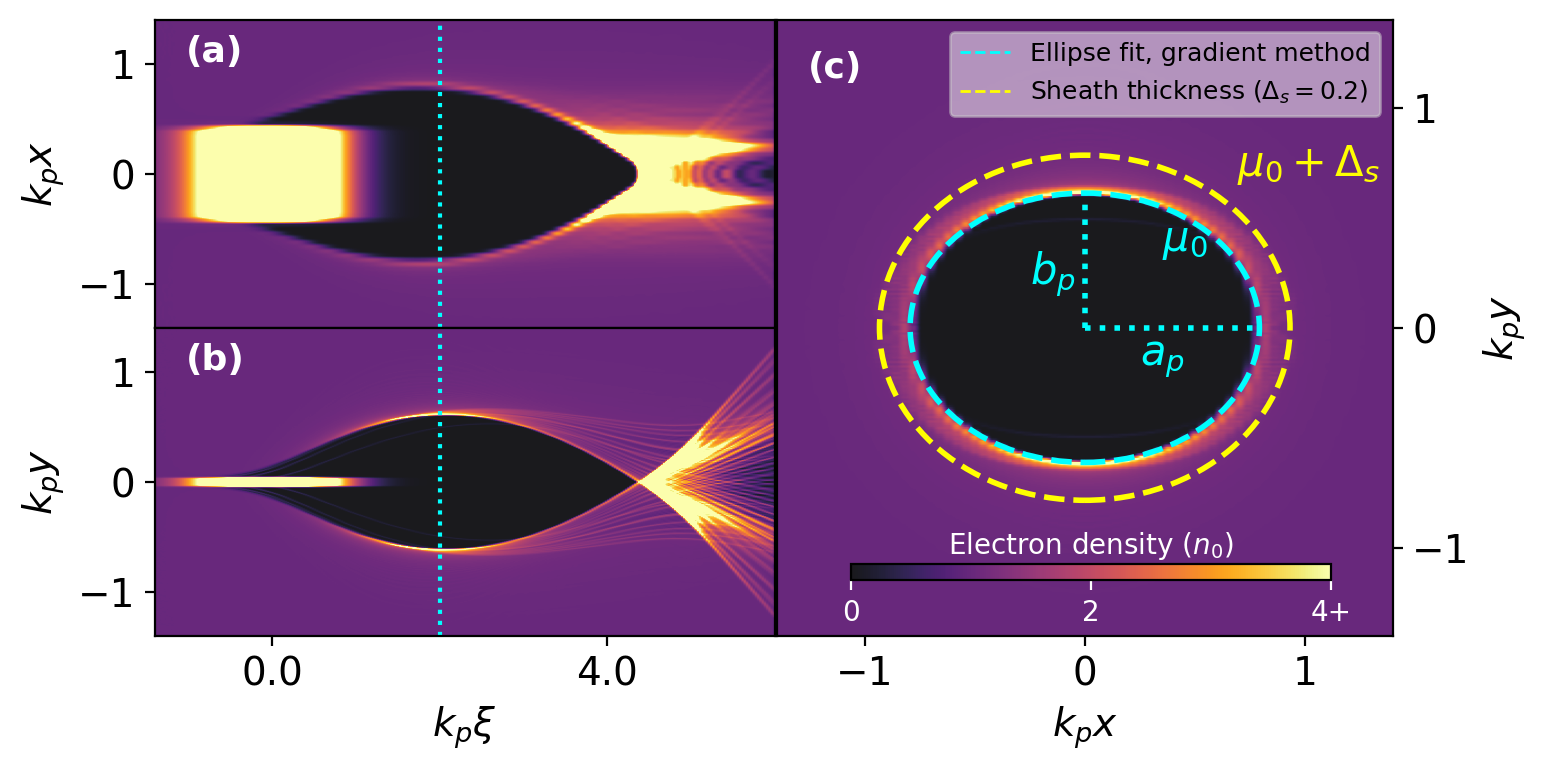}
    \caption{Plasma wakefield created by a uniform current driver with beam density $n_b = 15$, with spot sizes: $a$ = 0.424, $b$ = 0.0424. Longitudinal slices are shown in (a) the \textit{x-z} plane and (b) the \textit{y-z} plane, as well as (c) the transverse slice displaying the elliptical profile.}
    \label{fig:asymmetry}
\end{figure}

Note that the forces on the beam in the ultra-relativistic limit $v_b\rightarrow c$ are all simply derived from $\psi$. Here we concentrate on the plasma response instead of the beam When the driver interacts with an underdense plasma, plasma electrons are strongly repelled by the Coulomb force due to the dense beam charge, with magnetic effects becoming important for the relativistic plasma response found in the blowout regime. This repulsion leads to non-laminar plasma motion, which leads to evacuation of the plasma electrons from the beam channel and formation of a blowout sheath surrounding a plasma-electron-free cavity.

In order to display the most general plasma response, we examine here a short-beam case, corresponding to the excitation of large longitudinal wakes, and the dynamic evolution of the bubble dimensions (see Figure \ref{fig:asymmetry}). We will demonstrate that this scenario can be effectively explained using our theoretical model. To further advance our understanding of the interplay between plasma response and beam dynamics, we subsequently concentrate on a long-beam analysis. This is a particular, challenging scenario, and obtaining of a theoretical framework,  yields useful results, such as beam focal-matching conditions which are necessary to avoid  emittance growth during beam transport \cite{tjmehrling_2012}.

When considering an elliptically-shaped drive beam, the cross-sections of the blowout cavity and its accompanying boundary sheath also take on elliptical forms. The asymmetry of the wakefield produced by the elliptically-shaped beam are shown in the longitudinal slices displayed in Fig. \ref{fig:asymmetry}. The asymmetry drive more complex sheath electron trajectories after the bubble than in the symmetric case, where highly local trajectory crossing results in a large near-axis density spike. Here this feature is missing. The transverse  slice of the plasma shows an elliptical cross-section created by the evacuated plasma electrons, measured by semi-axes $a_p$ and $b_p$ and accompanied by the thin, dense plasma electron sheath of width $\Delta_\rho$. There is also a plasma return current $J_z $ outside the bubble having thickness $\sim k_p^{-1}$, \textit{i.e.} $\Delta_j\approx 1$. For the simplicity, we treat a full source sheath thickness, $\Delta_s$, taken to be equal to $\Delta_\rho$, as $\rho-J_z \approx \rho$ \cite{yi_2013} ($J_z $ has associated $v_z\ll c$).  For simplicity we assume a uniform sheath density distribution that does not vary with angular direction and is confocal with the ion column ellipse. The simplifying confocal condition stems from our choice of distribution. While this can be extended to other  distributions more reflective of simulations, we show in the Appendix that it yields a wake potential solution that vanishes outside the sheath, per the assumption that EM fields are not present outside of the sheath. 

Integrate the charge continuity relation over the transverse plane, with the divergence theorem we obtain the conservation of $S$ in each $\xi$-slice, $\dv{}{\xi}\int S dA=0$. With the absence of any source ahead of the driver, this conservation results in net zero source on each transverse plane, $\int S dA = 0$, permitting the sheath density to be found. 

We next introduce elliptical coordinates $(\mu, \nu)$ using the substitution $x=c_p \cosh{\mu} \cos{\nu}$ and $y = c_p \sinh{\mu} \sin{\nu}$, with $\mu_0$ and $c_p=\sqrt{a_p^2-b_p^2}$ defining the elliptical blowout boundary and focal length of the ellipse, respectively. The sheath is confocal to the ellipse defined by $\mu_0$. We can now construct the Poisson equation with defined source terms for each transverse slice:
\begin{equation}
\begin{split}
    -\nabla_\perp^2 \psi &= -\frac{2}{c_p^2 (\cosh{2\mu} - \cos{2\nu})} \left( \frac{\partial^2 \psi}{\partial \mu^2} + \frac{\partial^2 \psi}{\partial \nu^2} \right)\\
    & = 
\begin{cases}
    1 & \text{$\mu<\mu_0$, } \\
    -\frac{\sinh2\mu_0}{\sinh(2\mu_0+2\Delta_s)-\sinh2\mu_0} & \text{$\mu_0<\mu<\mu_0+\Delta_s$, } \\
    0 & \text{$\mu>\mu_0+\Delta_s$.}
\end{cases}
\end{split}
\label{eq:newpoisson}
\end{equation}
Here, we assume that $\Delta_s$ is small compared to $\mu_0$. We are interested in $\psi$ inside the ellipse, and so we can expand the equation to first order in $\Delta_s$:
\begin{equation}
\begin{split}
    \psi|_{\mu<\mu_0} &= -\frac{c_p^2}{8}\Biggl[ \cosh2\mu  - \cosh2\mu_0 + \left(1 - \frac{\cosh2\mu}{\cosh2\mu_0}\right)\cos2\nu\\
    &+ \Delta_s\left(\sinh2\mu_0 - \tanh2\mu_0\frac{\cosh2\mu}{\cosh2\mu_0}\cos2\nu\right)\Biggr]
\end{split}
\label{eq:wake}
\end{equation}
 Converting the results to Cartesian coordinates using $x = c \cosh{\mu} \cos{\nu}$ and $y = c \sinh{\mu} \sin{\nu}$, we find
\begin{equation}
\begin{split}
    \psi(x,y) &= -\frac{x^2 b_p^2 + y^2 a_p^2 - p_p^4}{2d_p^2}\\
    &- \frac{\Delta_s p_p^2}{2 d_p^4} \Biggl[(x^2 - y^2)c_p^2 - a_p^4 - b_p^4\Biggr],
\end{split}
\label{eq:wakepotential_cart}
\end{equation}
where $p_p^2 = a_p b_p$ and $d_p^2 = a_p^2 + b_p^2$. The potential $\psi$ which dictates  the beam electron motion  is quadratic in both $x$ and $y$ in the blowout cavity. As $\psi$ has the form of a two-dimensional harmonic oscillator, focal characteristics and matched (equilibrium propagation) beam conditions in both transverse axes are derivable from Eq. \ref{eq:wakepotential_cart}. In general, the wakefields can be derived from the gradient of the wake potential, as:
\begin{equation}
 \begin{split}
    W_{x} &= E_x - B_y = -\pdv[] {\psi}{x} = x \Biggl[\frac{b_p^2}{d_p^2} + \Delta_s \frac{p_p^2 c_p^2}{d_p^4}\Biggr] \\
    W_{y} &= E_y + B_x = -\pdv[] {\psi}{y} = y \Biggl[\frac{a_p^2}{d_p^2} - \Delta_s \frac{p_p^2 c_p^2}{d_p^4}\Biggr]\\
    W_z &= E_{z,p} = \frac{\partial \psi}{\partial \xi} = \frac{p_p^2}{d_p^4}\Big[C_1(x^2 - y^2) + C_2\Big]\\
    &+ \frac{\Delta_s}{2d_p^6}\Bigl[C_3(x^2 - y^2) + C_4\Bigr]
 \end{split}
 \label{eq:fields}
 \end{equation}
 where $C_1 = a_p'b_p - b_p'a_p$, $C_2 = a_p'b_p^3 + b_p'a_p^3$, $C_3 = C_1(a_p^4 - 6a_p^2b_p^2 + b_p^4)$, $C_4 =(a_p^6 + b_p^6)(p_p^2)' + (a_p^5)'b_p^3 - a_p^5(b_p^3)' + a_p^3(b_p^5)' - (a_p^3)'b_p^5$ and $'$ indicates $\partial/\partial\xi$. We estimate the sheath thickness in the transverse axes by reverting to Cartesian coordinates, \textit{i.e.} $\Delta_x \approx b_p \Delta_s$ and $\Delta_y \approx a_p \Delta_s$. 
 
 Simulations indicate that a thin sheath, $\Delta_s < 0.2$, yields  accurate results for $\vec{W}_\perp$ and we extend this result to the simplifying approximation ${\Delta_s\rightarrow0}$.   We then proceed without using $\Delta_s $ as a fitting parameter, as was done in Ref. \cite{golovanov_2023}. Additionally, as we assume that no EM fields exist outside the cavity due to the sheath shielding,  we set $\psi$ to be zero everywhere outside the sheath, giving $\nabla_\perp^2 \psi(\xi) = -1$ inside the bubble, with boundary condition  $\psi \vert_{ \partial\Omega(\xi)} = 0$.  This recovers the previous result without reliance on ${\Delta_s}$: $\psi=-(x^2 b_p^2 + y^2 a_p^2 - p_p^4)/2d_p^2$. With axisymmetry, $a_p = b_p$, we attain the familiar expression $\psi = -\left(r^2-r_b^2\right)/4$, where $r$ is the radial coordinate and $r_b$ the blowout radius. 
 
\begin{figure}
    \centering
    \includegraphics[width=\columnwidth]{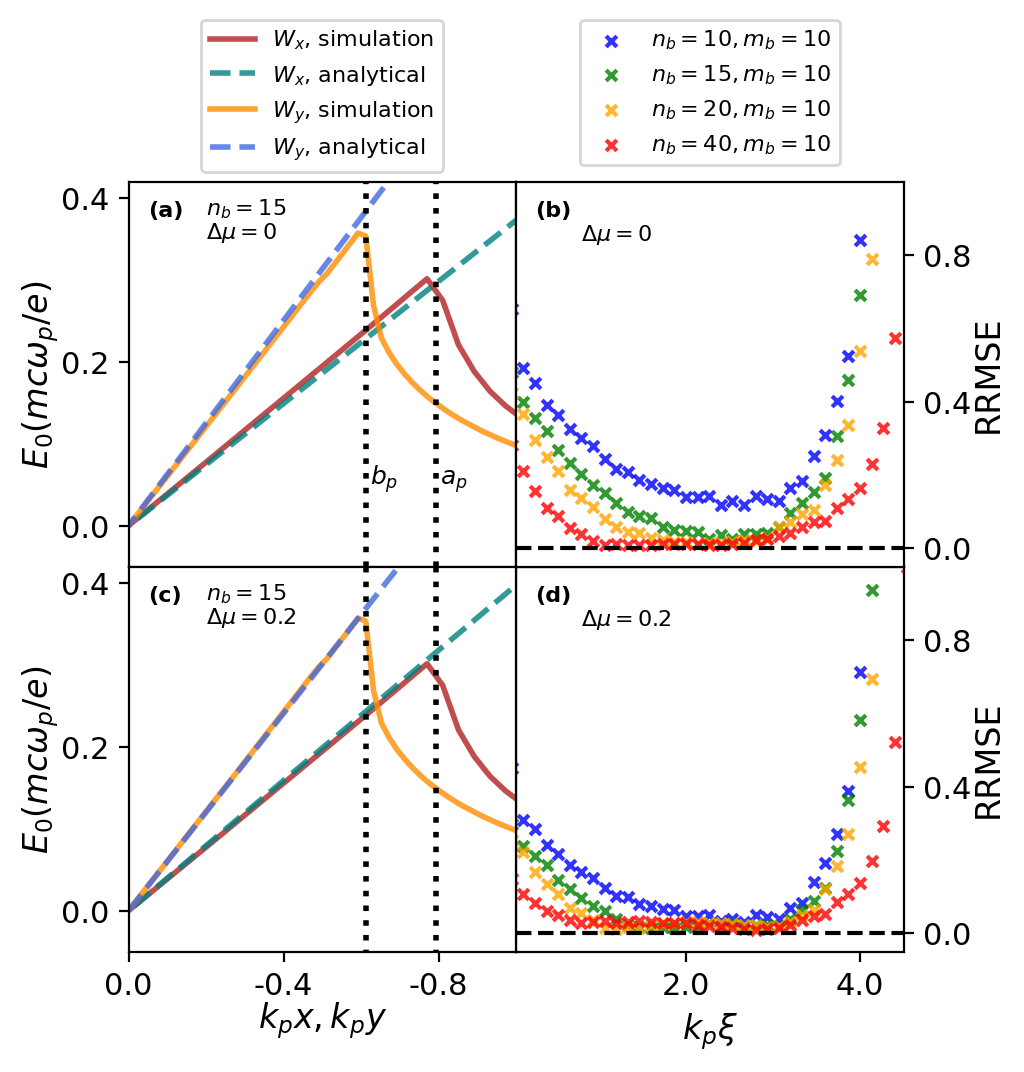}
    \caption{Short beam ($\sigma_z = 0.5$) driver, corresponding to the case of Fig. \ref{fig:asymmetry}. (a) Transverse wakefield line-outs of the wake. (b) Comparison of the analytical transverse wakefield $W_\perp$ from different $n_b$, calculated using the fitted blowout boundary. (c)-(d) show the same analysis with the inclusion of a finite sheath ($\Delta \mu=0.2$). }
    \label{fig:shortbeam}
\end{figure}

 We validate our approach to the general flat beam-plasma interaction by comparing the output of the PIC simulations with our analytical results. The axes of the resulting ellipse are found by numerically evaluating the boundary positions via a least-squares fitting of boundary points marking the position of maximum density gradient for 100 radial line searches taken at uniformly spaced angles. The relative root mean square error (RRMSE) between the PIC-simulated transverse fields and the fields predicted by Eq.~\ref{eq:fields} with the geometry found by the elliptical fit to the boundary points is shown in Figure \ref {fig:shortbeam}. The RRMSE is calculated by comparing the analytical and numerical fields at each data point within the blowout boundary, The elliptical model's predictions for transverse wakes closely match simulations.

 While it is not the main thrust of the present work, we add here some relevant comments on the longitudinal wake $W_z$. The Panofsky-Wenzel theorem \cite{panofsky_1956} as applied to plasma wakefields gives $\bm{\nabla}_\perp W_z = \partial_\xi W_\perp$., a result that follows due to the conditon that $\vec{W}=-\bm{\nabla}\psi$. In a symmetric blowout with $v_\phi=v_b=c$, $W_z$ is thus independent of transverse coordinate. It is also independent of the sheath thickness in the axisymmetric case. However, with an elliptical bubble shape, $W_z$ takes on quadratic $x$- and  $y$-dependence. Further, it varies with the sheath thickness  and the shape of the $\rho$ and $J_z$ distributions. This makes it challenging to estimate the $\xi$-dependence on $W_z$ as errors in determining $a_p'$, $b_p'$ and $\Delta_s'$ accumulate. However, the key result that the quadratic coordinate dependence is clear from our results (refer to Supplemental Material for comparison with simulations \cite{supplemental}). We additionally note that the $C_1$ term in equation \ref{eq:fields}, is dependent on the longitudinal derivative of the ellipticity of the cavity, $\alpha_p=a_p/b_p$, and is small compared to the central response term, $C_2$. While this term leads to a small increase in slice energy spread, the  total energy spread mainly arises from the chirp induced by the wakefield $\xi$-dependence. As the witness beam is typically much smaller than the bubble, multi-pole effects of the accelerating field would be small.

To most straightforwardly exploit the utility of our analytical results, we proceed to calculate the location of the elliptical boundaries using the beam parameters in the long beam limit, $\sigma_z \gg 1$. This permits one to neglect the fields' variations in  $\xi$ in the beam region. We are thus searching for an equilibrium scenario, and we examine the force on the plasma electrons at a position $\bm{r}$ near the sheath. Considering the plasma electron's transverse velocity $\bm{v_{\perp}}=d \bm{r_{\perp}}/dt=\left(1-v_z\right) d \bm{r_{\perp}}/{d \xi} = \gamma^{-1} (1+\psi) d \bm{r_{\perp}}/{d \xi}$ in combination with Eq. \ref{eq:energy}, 
\begin{equation}
 \bm{F_{\perp}}=\left(1-v_z\right) \frac{d \bm{p_{\perp}}}{d \xi}=\frac{1}{\bar{\gamma}}(1+\psi) \frac{d}{d \xi} \left((1+\psi) \frac{d}{d \xi} \bm{r_{\perp}}\right)
\end{equation}
\begin{equation}
\bm{F_{\perp}}\vert_{ \partial\Omega} = \frac{1}{\bar{\gamma}}\left(\frac{d \psi}{d \xi} \frac{d \bm{r_{\perp}}}{d \xi}+\frac{d^2 \bm{r_{\perp}}}{d \xi^2}\right)\bigg\vert_{ \partial\Omega}=0,
\label{eq:equate_transverse}
\end{equation}
where $\partial{\Omega}$ indicates the sheath boundary condition. Neglecting possible variations in $\xi$ allows us to balance the transverse forces at the boundaries. in this regard, we also assume that the sheath electron longitudinal velocity $v_z$ does not depend on transverse direction. This aids solution of the above equations by providing a linear relationship between $\phi_e$ and $A_e$, where the role of  $v_z$ enters as as a small correction. 
Using Eq. \ref{eq:maxwell} to convert to a potential description and using $v_b=1$, we have

\begin{equation}
\begin{split}
    \bm{F_\perp} &= \bm{\nabla}_\perp \phi_i + \bm{\nabla}_\perp \phi_e+ \bm{\nabla}_\perp \phi_b \\ + &\partial_\xi \bm{A_\perp} - (\bm{\nabla_\perp}\bm{A})\cdot\bm{v} + (\bm{v}\cdot\bm{\nabla})\bm{A}_\perp
\end{split}
\end{equation}
Using Eq. \ref{eq:equate_transverse} and neglecting the longitudinal variation of the fields and the transverse velocity, leads to the relation
\begin{figure}
    \centering
    \includegraphics[width=\columnwidth]{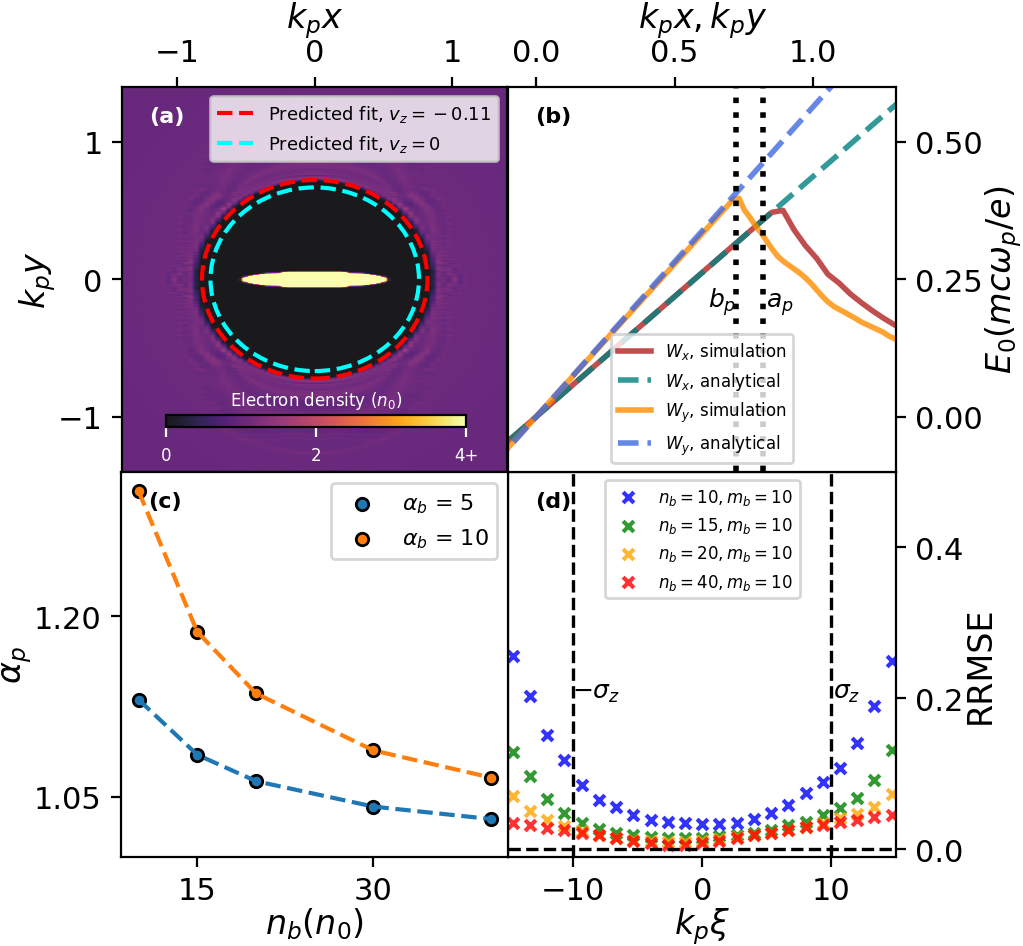}
    \caption{Long beam ($\sigma_z = 10$) driver: (a) Analytical calculation for the blowout shape using a beam with $n_b = 20$, $a = 0.5$ and $b = 0.05$. (b) Transverse wakefield lineouts of the wake, calculated using the predicted blowout boundary. (c) predicted blowout ellipticity at the center of wake vs beam density and beam ellipticity. (d) Comparison between the analytically calculated transverse wakefield and the simulation results.}
    \label{fig:longbeam}
\end{figure}

\begin{equation}
\begin{split}
    (1+v_z) \bm{\nabla}_{\perp} \psi|_{\partial\Omega}-v_z \bm{\nabla}_{\perp} \phi_i|_{\partial\Omega}+(1-v_z) \bm{\nabla}_{\perp} \phi_b|_{\partial\Omega}
    = 0 .
\end{split}
\end{equation}

Since the cavity is effectively evacuated of plasma electrons, we can obtain the scalar potential due to the ions alone, $\phi_i(\xi)= -\frac{x^2 b_p(\xi) + y^2 a_p(\xi)}{2(a_p(\xi) + b_p(\xi))}$ \cite{ellipse_shubaly_1975}. We first neglect $v_z$ to find the zero-th order equation in the electrostatic limit. Using Ampere's law, $\oint \bm{B}\cdot\bm{dl} = \int \left(\bm{J} + \epsilon_0 \pdv{\bm{E}}{t} \right) \cdot \bm{da}$, we integrate over the transverse plane to remove the left hand side term, finding $I_{z,beam} + I_{z,elec} + \int^\infty_0 \pdv[2]{\psi}{\xi} da = 0$. In the long beam limit, the integral term is negligible. Consequently, the beam current and the total plasma return current are nearly equal. With the return current located within a layer of thickness of $\Delta_j=1$ and assuming $v_z$ is constant across this region, we find that $v_z = \frac{\lambda_b}{\pi(a_p+1)(b_p+1)}$, where $\lambda_b$ is the beam charge per unit length. This permits inclusion the EM features of the sheath, beyond the zero-th order analysis, for weak blowout scenarios. For strong blowouts, a detailed treatment of the sheath is challenging. Fortunately, in this limit an axisymmetric bubble is inherently approached and the point is moot.

We now use the electric fields of the elliptical drive beam \cite{Parzen_2001} to simultaneously find $\psi$ and the elliptical semi-axes $a_p$ and $b_p$ such that the transverse forces on the ellipse boundaries are minimized. We verify the results by calculating the blowout shape for a long beam driver, predicting the transverse fields and calculating the RRMSE values in Fig. \ref{fig:longbeam} (see Supplemental Material for the details on PIC simulations \cite{supplemental}). In this way, we demonstrate the capability of determining the elliptical boundaries, $a_p$ and $b_p$ for given beam parameters. In this way, we have a path to self-consistently determining the beam sizes and blowout dimensions. The beam sizes are dependent on the emittances and the linear focusing strengths in $x$ and $y$. . We write equations of motion for the transverse beam dynamics inside the asymmetric blowout cavity:

\begin{equation}
    x^{\prime \prime}(\xi)+K_{x} x(\xi) =0;\ y^{\prime \prime}(\xi)+K_{y}  y(\xi) = 0
\end{equation}

\begin{equation}
    K_x=K_r \frac{2}{1+\alpha_p^2};\ K_y=K_r \frac{2\alpha_p^2}{1+\alpha_p^2}
\end{equation}
where $K_r =1/2\gamma$ corresponds to the normalized linear focusing strength due to the ions in an axisymmetric ion column. We observe that the focusing is a superposition of monopole and quadrupole strengths, $K_r $ and $\pm K_r \frac{\alpha_p^2-1}{\alpha_p^2+1}$, respectively, as could be  deduced directly from the elliptical symmetry ion column. The  equilibrium propagation or \textit{matching} conditions for a beam in the asymmetric bubble are then given by:
\begin{equation}
    \sigma_{m,\eta} = \sqrt{\frac{\sqrt{K_{\eta}^{-1}}\epsilon_{n,\eta}}{\gamma}}\ \rightarrow \ \frac{\sigma_{m,x}}{\sigma_{m,y}} = \sqrt{\frac{\epsilon_{n,x}}{\epsilon_{n,y}}\alpha_p}\label{matchsig}
\end{equation}
where $\eta\in\left\{x,y\right\}$ and $\epsilon_{n,\eta}$ are the normalized emittances.  The matched beam aspect ratio is determined by both the emittance ratio and the wake ellipticity. Equation \ref{matchsig} can be used to match the beam inside a given plasma profile through a simple iterative procedure we will describe in a future paper. 

It is useful to examine limits.  It can be seen that for strong blowouts where $\alpha_p\simeq 1$, the beam's equilibrium sizes can be easily deduced from the familiar axisymmetric ion-column focusing conditions. In the weak blowout limit, the bubble boundary closely follows the beam contours ($\alpha_p\simeq \sigma_{m,x}/\sigma_{m,y}$), and one may deduce that $ \sigma_{m,x}/\sigma_{m,y}\simeq \epsilon_{n,x}/\epsilon_{n,y}$. In weak blowout, the beam asymmetry is further enhanced by the associated focal asymmetry.

In this paper, we have created a phenomenological model for  the plasma structures formed by an elliptical beam's wake. We have concentrated on the crucial issue of the focusing forces in this scenario, which imply significant changes in the beam dynamics compared to the axisymmetric case. Unlike the  axisymmetric case, where the transverse wake forces remain uniform along the beam axis, the asymmetric case has interesting and useful properties, as  the elliptical boundaries varying in $\xi$ result in $\xi$-dependent, but still linear, focusing forces. This variation results in decoherence of the head-to-tail oscillations of the beam particles,   mitigating the effects of the hosing instability \cite{BNS_1983,Whittum_1991_hosing,mehrling_2018}. Asymmetric drivers offer another key advantage in creating a disparity in focusing forces between the transverse axes, detuning the resonant emittance mixing of a  flat accelerating beam in the presence of nonlinearities \cite{deiderichs_2024}, as the resonance condition $k_{\beta,x} \simeq k_{\beta,y}$ is violated. Further, the linearity of the focusing forces enables matching, even slice-by-slice, \cite{manwani:ipac24-mopr64}, resulting in suppressed emittance growth due to the transverse coordinate mismatch effects. These insights are  essential for present plasma wakefield experiments and future plans for plasma afterburner colliders using asymmetric beams. Such experimental scenarios include an experiment developed for the Argonne Wakefield Accelerator for examining the propagation of a particle beam in a plasma with asymmetric transverse emittances ($\epsilon_x/\epsilon_y \approx 100, \sigma_z \approx 1.5$) \cite{Xu_2019,manwani_ipac_21,yadav}, as well as the development of an asymmetric underdense plasma lens at FACET-II  induced by using an asymmetric beam ($\sigma_x/\sigma_y \approx 10, \sigma_z \approx 1$) at the plasma \cite{clarke_2022_lens_result}. The theory of wakefields in an elliptical cavity approach can also be utilized in the context of laser wakefield acceleration (LWFA) \cite{tajima_1979,WeiLu_3D,Rao_2020} where an elliptical blowout cavity is created. 

While this Letter lays the foundation for understanding key aspects of a asymmetric wakefields driven by elliptically-shaped beams, further research is necessary.  To aid in experimental studies, a comparison of the asymmetric blowout bubbles driven by Gaussian and flat-top beams is presented in Ref. \cite{kang_2024}. These experimental scenarios will include to another long beam transverse matching application , that of a flat-beam adiabatic plasma lens demonstration. Further, we must widen the applicability of the theory to the short-beam regime where acceleration is dominant. Efforts towards this goal here are underway, as seen by the additional results given in the Supplemental Material \cite{supplemental}. These illustrate steps towards understanding the longitudinal wakefield temporal dependence.  We are currently working on a generalization of previous  axisymmetric analyses \cite{weilu_2006, golovanov_2023} that exploits the established ellipsoidal shapes to give the potentials and fields in the bubble region behind the drive beam.  This work will provide a strong complement to the results presented here. 

The authors would like to thank Nathan Majernik for insightful discussions. This work was performed with the support of the US Dept. of Energy under Contract No. DE-SC0017648 and DE0SC0009914. This work used resources of the National Energy Research Scientific Computing Center, a US DOE Office of Science User Facility, operated under Contract No. DE-AC02-05CH11231.

\section{Appendix}
\renewcommand{\theequation}{A\arabic{equation}}
\renewcommand{\thefigure}{A\arabic{figure}}
\setcounter{equation}{0}
\setcounter{figure}{0}
We start with the analysis done by Regenstreif to find the potential and field produced by a uniform beam with charge density $\rho$ inside a confocal vacuum chamber\cite{Regenstreif_1977}. We define two boundaries: one for the vacuum chamber and one for the beam, respectively $\mu = \mu_1$ and $\mu = \mu_0$.

\begin{equation}
\begin{split}
\psi_{\mu>\mu_0} &= \frac{\rho c_p^2}{8 } \left[ 2(\mu_1 - \mu) - \frac{\sinh{2(\mu_1 - \mu)}}{\cosh{2\mu_1}} \cos2{\nu} \right] \sinh{2\mu_0} \\
\psi_{\mu<\mu_0} &= \frac{\rho c_p^2}{8 } \Bigg(\left[ \frac{\cosh{2(\mu_1 - \mu_0)}}{\cosh{2\mu_1}} \cosh{2\mu} - 1 \right] \cos{2\nu} \\
&+ \cosh{2\mu_0} - \cosh{2\mu} + 2\left( \mu_1 - \mu_0 \right)\sinh{2\mu_0}\Bigg)
\end{split}
\end{equation}
Taking the chamber to be at infinity, $\mu_1\rightarrow\infty$, we have the substitutions $    \lim_{\mu_1\to\infty}\frac{\sinh(2\mu_1 - 2\mu)}{\cosh2\mu_1} = \cosh2\mu - \sinh2\mu$ and $\lim_{\mu_1\to\infty}\frac{\cosh(2\mu_1 - 2\mu_0)}{\cosh2\mu_1} = \cosh2\mu_0 - \sinh2\mu_0$. Ignoring the constant term that contributes to infinity, a solution for the isolated beam can be obtained:

\begin{equation}
\begin{split}
\psi_{\mu>\mu_0} &= -\frac{\rho c_p^2}{8 } \left[2\mu + (\cosh2\mu - \sinh2\mu) \cos2\nu \right] \sinh2\mu_0 \\
\psi_{\mu<\mu_0} &= -\frac{\rho c_p^2}{8 } \Big([1-(\cosh2\mu_0 - \sinh2\mu_0) \cosh2\mu] \cos2\nu \\
&+ \cosh2\mu_0 - \cosh2\mu -2\mu_0\sinh2\mu_0\Big)
\end{split}
\end{equation}
We can use this analysis, to construct the source term, $S=\rho - J_z$ by using two confocal ellipses overlapping each other; $\mu_{in}=\mu_0$ and $\mu_{out} = \mu_0 + \Delta_s$. To yield the desired source distribution defined in the paper, we set the source densities to be $S_{in} = - S_{\Delta_s}$ and $S_{out} = \rho_i + S_{\Delta_s}$. Applying the conservation of $S$ in the transverse plane,  $\int^{2\pi}_{0}\int^{\mu_{out}}_{0}(S_{in} + S_{out})\frac{c_p^2}{2} (\cosh 2\mu - \cos 2\nu)d\mu d\nu = 0$, yielding:
\begin{equation}
    S_{\Delta_s} = -\frac{\rho_i \sinh{(\mu_0 + \Delta_s)}}{\sinh{(\mu_0 + \Delta_s)}-\sinh{\mu_0}}    
\end{equation}

We now have the complete form of the source density and can find the $\psi$ using the superposition of the wake potentials of the two confocal ellipses. The combination of the conservation of the source term and the confocality condition has a key property: outside the sheath, the wakefield due to the sheath cancels out the wakefield due to the ion column, i.e $\psi|_{\mu>\mu_0} = 0$. The confocal model works well to capture this essential quality of the blowout regime, showing the elliptical symmetry of the fields. We are also interested in the form of $\psi$ inside the ion column to calculate the wakefields, and inside the sheath to calculate the electron trajectories. This can be calculated from the isolated beam solutions. 

\begin{figure}
    \centering
    \includegraphics[width=1\linewidth]{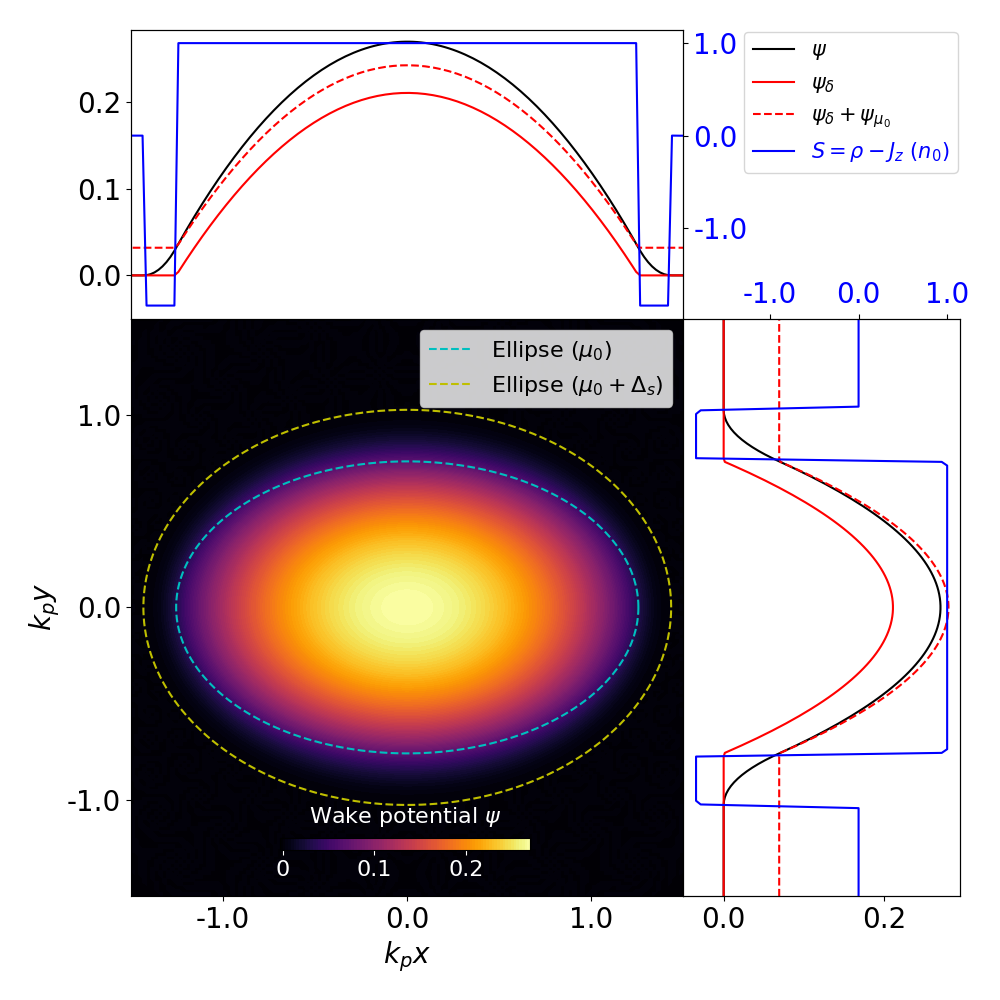}
    \caption{Wake potential calculated for a specific set of parameters  ($c_p=1$, $\mu_0=0.7$, $\Delta_s=0.2$). The plot illustrates the shape of the wake potential $\psi$ with the corresponding line-outs along the x and y axes. It also includes the line-outs of the source term $S$ and the delta-function potential $\psi_\delta$. }
    \label{fig:psi}
\end{figure}

Here, we simplify the former by introducing the assumption that the sheath thickness is small compared to the ion column ($\Delta_s/\mu_0<1$), yielding the solutions (up to first order in $\Delta_s$):

\begin{equation}
\begin{split}
    \psi|_{\mu<\mu_0 }&= -\frac{c_p^2}{8}\Biggl[ \cosh2\mu  - \cosh2\mu_0 + \left(1 - \frac{\cosh2\mu}{\cosh2\mu_0}\right)\cos2\nu \\
    &+ \Delta_s\left(\sinh2\mu_0 - \tanh2\mu_0\frac{\cosh2\mu}{\cosh2\mu_0}\cos2\nu\right)\Biggr]
\end{split}
\end{equation}

If we assume the sheath to be infinitesimal (${\Delta_s\rightarrow0}$), there is an alternative method that can be used to arrive at the same solution. Assuming no electromagnetic fields exist outside the blowout due to the shielding provided by the sheath at $\partial\Omega(\xi)$ (Dirichlet boundary condition), we may then set the wake potential to be zero everywhere outside the blowout region. Finally, the constant charge density inside the cavity from the ions alone enables us to solve for the wake. In this case, we have: $\nabla_\perp^2 \psi_\delta(\xi) = -1$, with  $\psi_\delta \vert_{ \partial\Omega(\xi)} = 0$. Then the particular solution is:
\begin{equation}
\psi_{\delta,p}=-\frac{a^2}{8}\left(\cosh (2 \mu)-\cosh \left(2 \mu_0 \right)+\cos (2 v)\right)
\end{equation}
We add a homogeneous solution such that potential is 0 at $\mu=\mu_0$:
\begin{equation}
\psi_{\delta,h}=\frac{a^2}{8}\left(\frac{\cosh (2 \mu)}{\cosh \left(2 \mu_0\right)}\right) \cos (2 v)
\end{equation}
Adding the particular and homogenous solutions gives us the same result as $\lim_{\Delta_s\to0}\psi$ =$\psi_\delta$.
\begin{equation}
\begin{split}
    \psi_\delta &= -\frac{c_p^2}{8}\left(\cosh{2\mu} - \cosh{2 \mu_0}+\left(1 - \frac{\cosh{2\mu}}{\cosh{2\mu_0}}\right)\cos{2\nu}\right) 
\end{split}
\label{eq:wakepotential_ellip}
\end{equation}

The exact wake potential solution including the sheath, along with the solution from the delta function approximation is shown in Fig. \ref{fig:psi}. Adding an offset to the delta function approximation allows us to study the differences between the two wake potential profiles. This comparison is crucial for analyzing the gradients, which are of primary interest.

\bibliography{aps}

\end{document}


\maketitle
\tableofcontents

\section{Ellipse fitting method}
The selection of data points is tuned to capture the structure of plasma electron distribution by choosing on region exhibiting the maximum gradient magnitude of plasma electron density. This is achieved by analyzing the transverse plasma density slice from a multitude of angles (total 100 angles range from $0$ to $2\pi$), ensuring a comprehensive angular coverage. For each specified angle, we identify and select point that is proximate to the maximal gradient of the plasma density along one angle. This approach allows us to ensure that, for every direction examined, the data points are selected and reflect the cavity's spatial structure. Upon selecting the optimal data points, our analysis progresses to determine the plasma's fitted elliptical shape. It relies on a direct application of least squares regression to the chosen points. By fitting these points to an elliptical model, the major axis and minor axis can be obtained to represent the overall shape of the blowout under the elliptical approximation. 

\section{Field minimization at boundary}

We point to Equation \ref{eq:equate_transverse} in the paper. In order to obtain the $a_p$ and $b_p$ that satisfies the above condition, a pair of $(a_p, b_p)$ is guessed, and along the elliptical shape, the value of $F_{x, y}$ and $E_{x, y}$ are taken. We have used the optimization methods to minimize the total sum below to obtain a numerical solution of $(a_p, b_p)$:

\begin{equation}
\begin{split}
    \min_{(a_p, b_p)} \sum_{x^2/a_p^2+y^2/b_p^2 = 1} \left((1+v_z) W_{x} - (1-v_z) E_{b, x} + v_z  E_{i, x}\right)^2 \\+ \left((1+v_z) W_{y} - (1-v_z) E_{b, y} + v_z  E_{i, y}\right)^2
\end{split}
\end{equation}

$v_z$ is treated as a constant in the process of optimization, adding weights to the wake fields and electric fields. The transverse electric field of the flat top beam in the x direction, is calculated using the formula given in [G. Parzen. (2001)]:
\begin{align}
    E_{x, b} = \frac{n_bx}{4(a^2 + t_1)^{1/2}((a^2 + t_1)^{1/2} + (b^2 + t_1)^{1/2})}
\end{align}

where, $t_1 = ((x^2 + y^2 - a_b^2 - b_b^2)^2/4 + (x^2b_b^2 + y^2a_b^2 - a_b^2 b_b^2))^{1/2} + (x^2 + y^2 - a_b^2 - b_b^2)/2$. The field in the y direction, $E_{y, b}$, follows similar form with a and b, x and y interchanged.

\section{PIC simulations : Parameters}

We ran PIC simulations using the OSIRIS code [R A Fonseca et al. Plasma Physics and Controlled Fusion, 50(12), 124034 (2008)] to obtain the simulated result for the cases where the beam is short ($\sigma_z< 1$)and long ($\sigma_z >> 1$) and compare them with our analytical prediction. The simulation results are normalized to plasma units such that length, time, and charge density are normalized to $k_p^{-1}$, $\omega_p^{-1}$, and $n_0$, respectively. The ions are set to be stationary while the driver is set to be not evolving in time. For the driver beam, we used a flattop beam where the beam density is constant inside of the elliptical beam shape in the transverse plane with a Gaussian density profile in the longitudinal direction.

\subsection{Short beam ($\sigma_z = 0.5$)}

For the short beam case, we used the simulation box size $3\times 3\times 8$ subdivided into 300, 150, and 600 grids along the $x$, $y$, and $z$ axes, with 2, 4, 1 particles per cell respectively. The time step is set to be $dt = 0.0025$, and the simulation is run for $t=30$. We use the following parameters for the driver beam: Beam ellipiticity $\alpha_p= 10$, Spot sizes: $a = 0.424$, $b =  0.0424$, and $\sigma_z = 0.5$. We scan through the beam density, $n_b = 10, 15, 20, 30, 40, 60, 80$, until the driver beam is strong enough to create a nearly axisymmetric blowout, to verify the analytical result. The figures below are at the center of the blowout:

\begin{figure}[H]
    \centering
    \includegraphics[width=\textwidth]{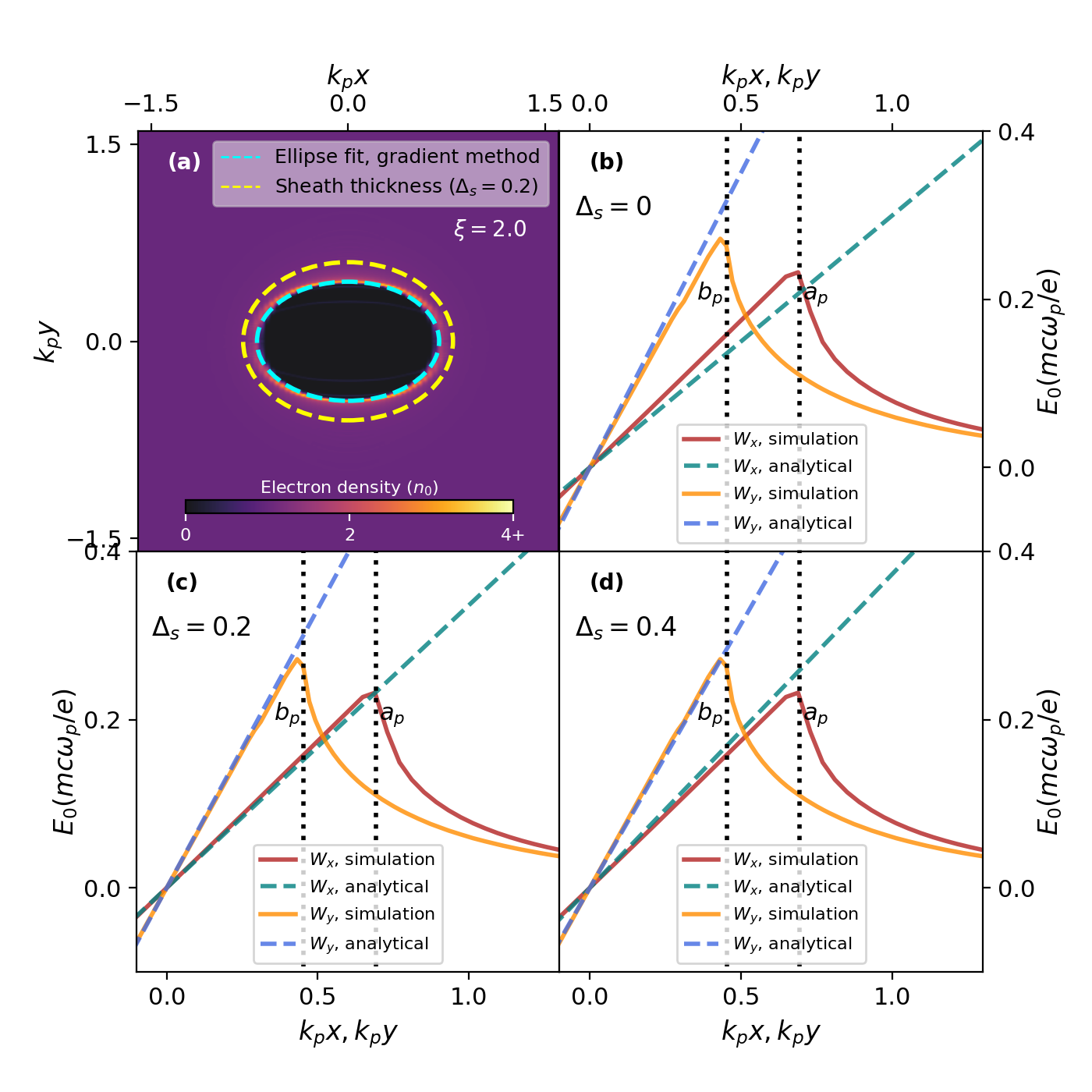}
    \caption{Short beam: Elliptical fit for the blowout shape using a beam with $n_b = 10$, $a =  0.424$, $b = 0.0424$ (a),  Transverse wakefield lineouts of the wake, calculated using the fitted blowout boundary with $\Delta_s = 0$(b), $\Delta_s = 0.2$(c), $\Delta_s = 0.4$(d)}
\end{figure}

\begin{figure}[H]
    \centering
    \includegraphics[width=\textwidth]{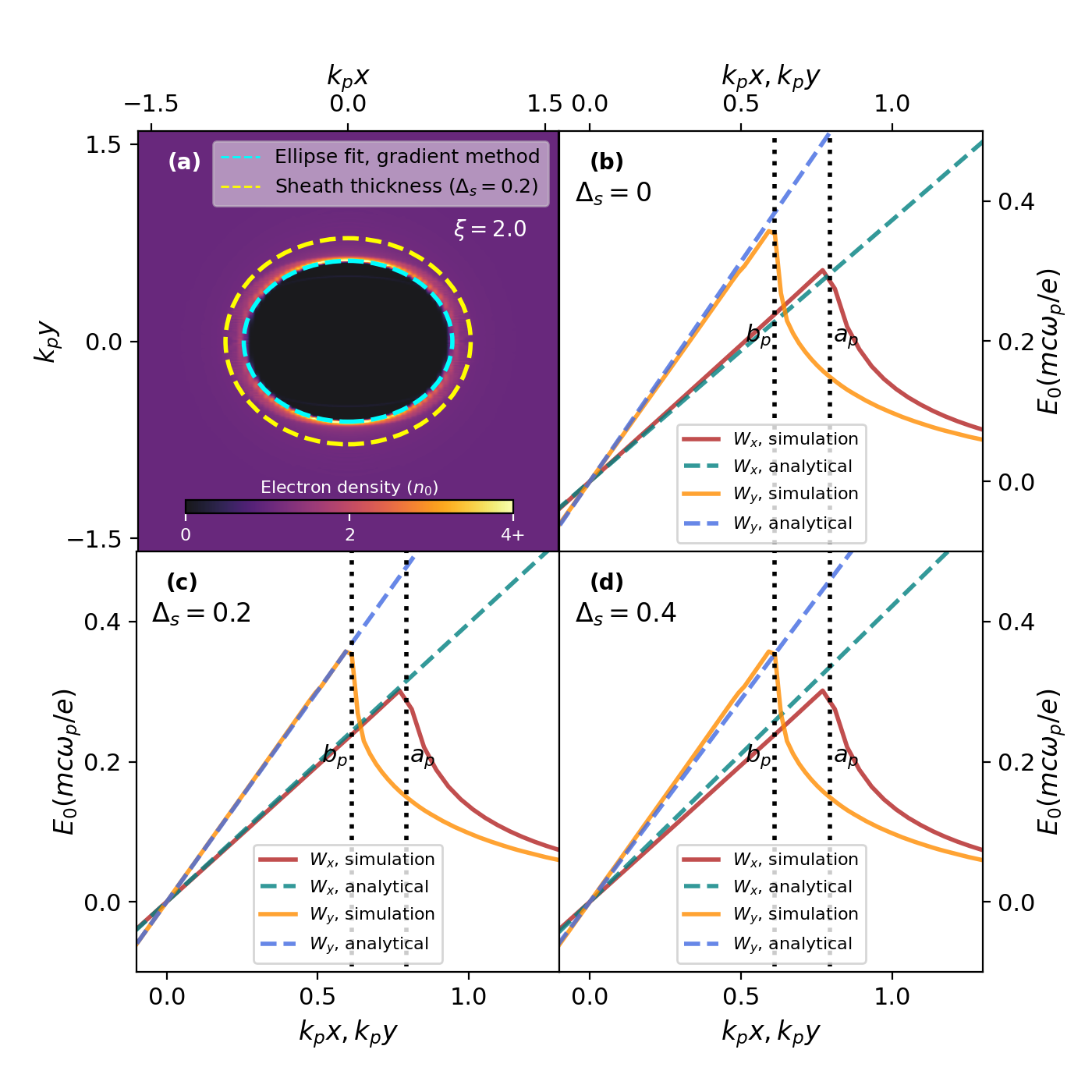}
    \caption{Short beam: Elliptical fit for the blowout shape using a beam with $n_b = 15$, $a =  0.424$, $b = 0.0424$ (a) Transverse wakefield lineouts of the wake, calculated using the fitted blowout boundary with $\Delta_s = 0$(b), $\Delta_s = 0.2$(c), $\Delta_s = 0.4$(d)}
\end{figure}

\begin{figure}[H]
    \centering
    \includegraphics[width=\textwidth]{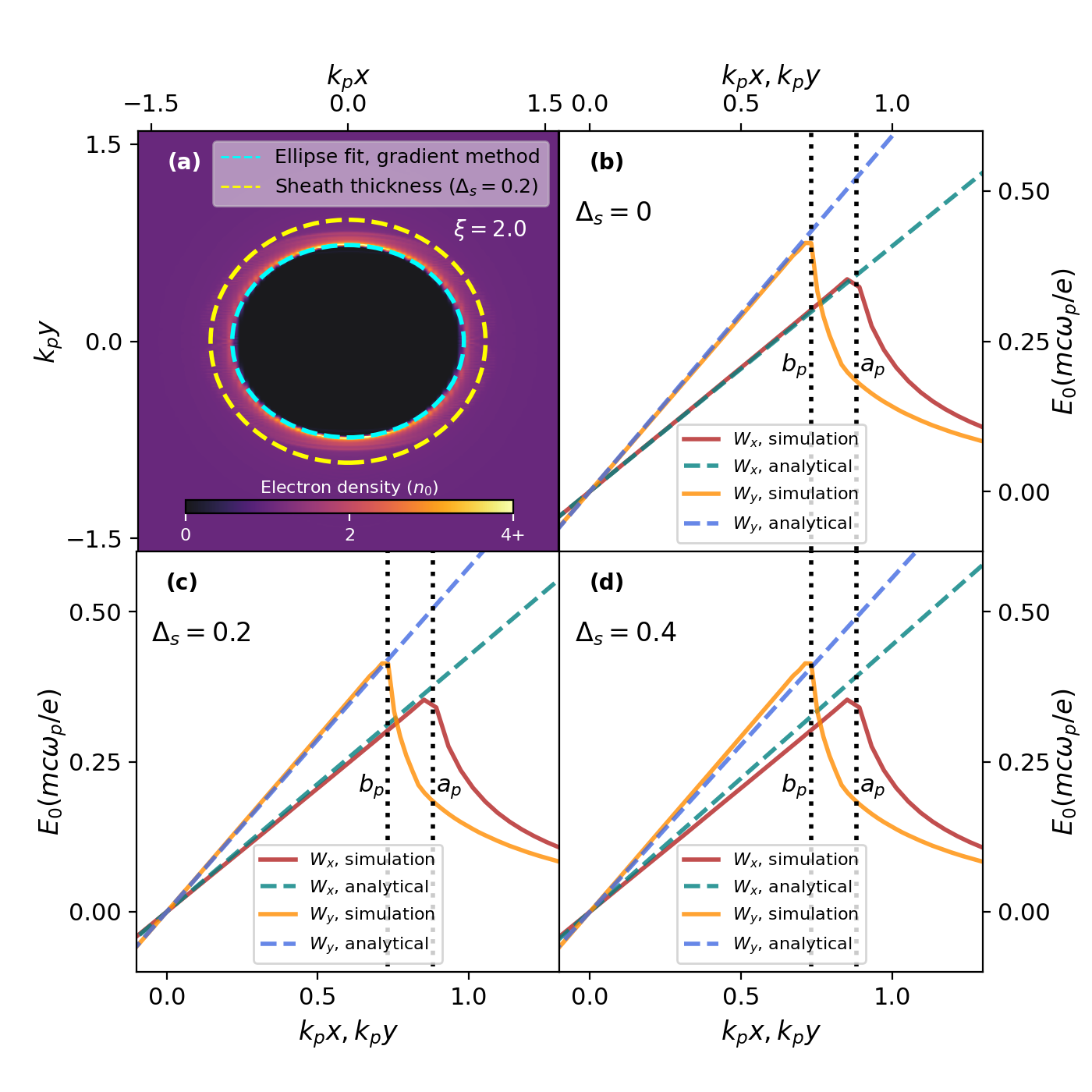}
    \caption{Short beam: Elliptical fit for the blowout shape using a beam with $n_b = 20$, $a =  0.424$, $b = 0.0424$ (a) Transverse wakefield lineouts of the wake, calculated using the fitted blowout boundary with $\Delta_s = 0$(b), $\Delta_s = 0.2$(c), $\Delta_s = 0.4$(d)}
\end{figure}

\begin{figure}[H]
    \centering
    \includegraphics[width=\textwidth]{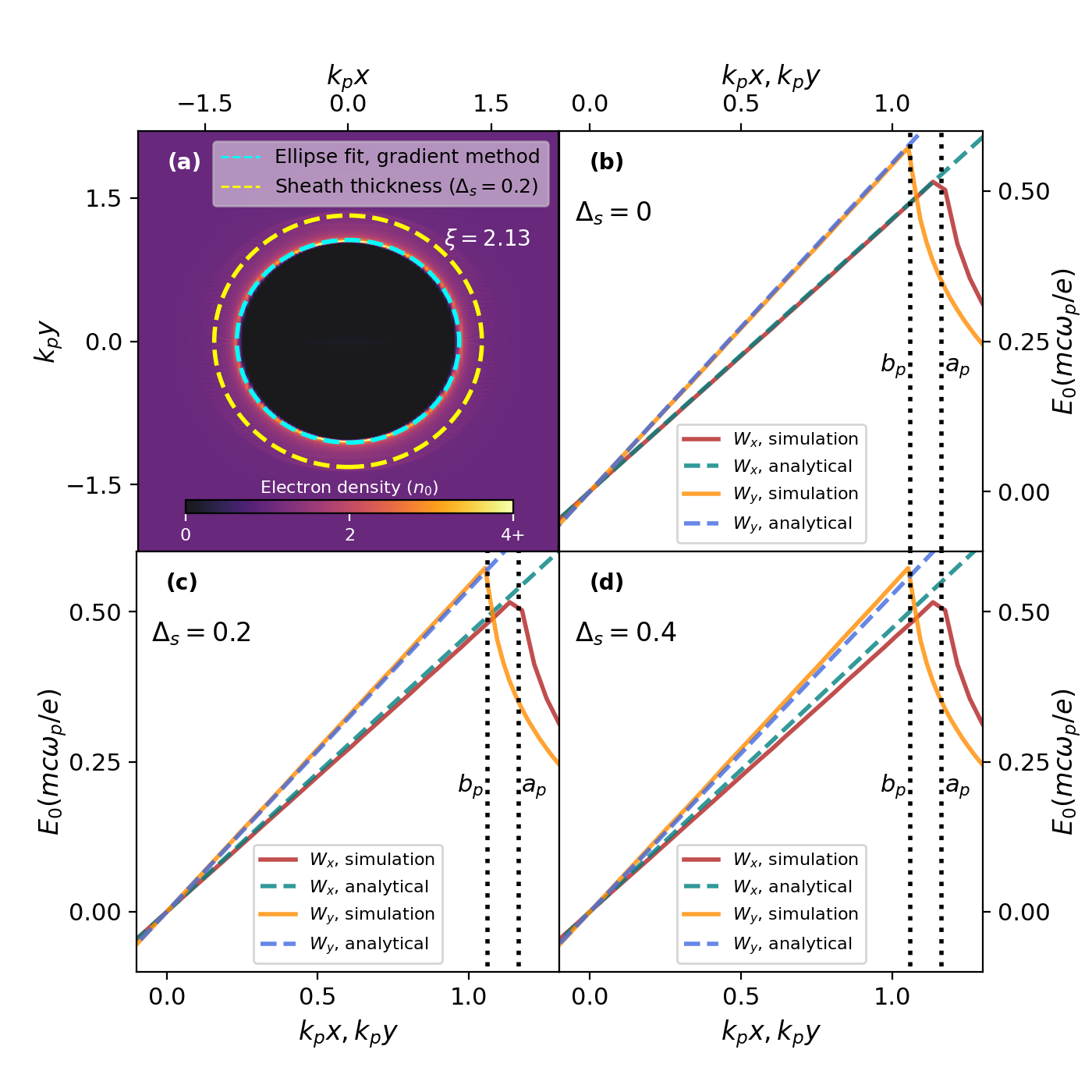}
    \caption{Short beam: Elliptical fit for the blowout shape using a beam with $n_b = 40$, $a =  0.424$, $b = 0.0424$ (a) Transverse wakefield lineouts of the wake, calculated using the fitted blowout boundary with $\Delta_s = 0$(b), $\Delta_s = 0.2$(c), $\Delta_s = 0.4$(d)}
\end{figure}


\subsection{Long beam ($\sigma_z = 10$)}
For the long beam case, we use the same beam parameters but change the longitudinal spot size, $\sigma_z = 10$. We used the simulation box size $6\times 6\times 80$ subdivided into 300, 150, and 600 grids along the $x$, $y$, and $z$ axes, with 2, 4, 1 particles per cell respectively. The time step is set to be $dt = 0.005$, and the simulation is run till $80$. We scan through different beam densities: $n_b = 10, 15, 20, 30, 40$. We have also tested for the case where the beam ellipticity is 5 ($a = 0.356$, $b = 0.0713$), where we used the simulation box size $6\times 6\times 80$ subdivided into 300, 150, and 300 grids along the $x$, $y$, and $z$ axes, with 2, 4, 1 particles per cell respectively. The time step is set to be $dt = 0.01$, and the simulation is run for $t=80$. For different beam ellipticities, we have ensured that the beam line density are the same to see the sole effect of driver's asymmetry on the blowout shape and wake field.

\begin{figure}[H]
    \centering
    \includegraphics[width=\textwidth]{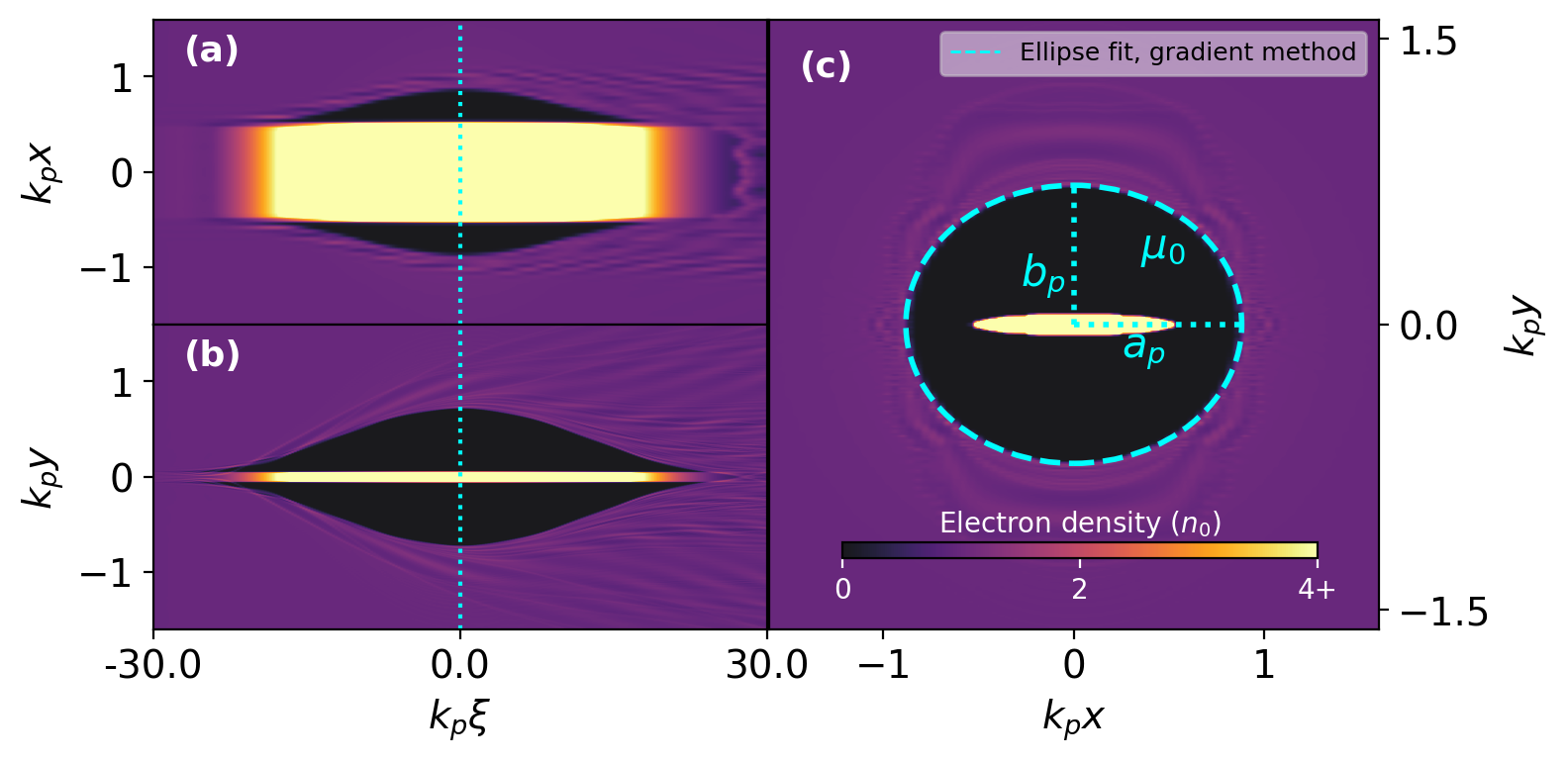}
    \caption{Plasma wakefield created by a uniform long beam ($\sigma_z = 10$) driver pulse with beam density $n_b = 20$, $a = 0.5$ and $b = 0.05$. Longitudinal slices are shown in (a) the \textit{x-z} plane and (b) the \textit{y-z} plane, as well as (c) the transverse slice displaying the elliptical profile ($\mu_0$ defines the blowout boundary in elliptical coordinate).}
\end{figure}

\begin{figure}[H]
    \centering
    \includegraphics[width=\textwidth]{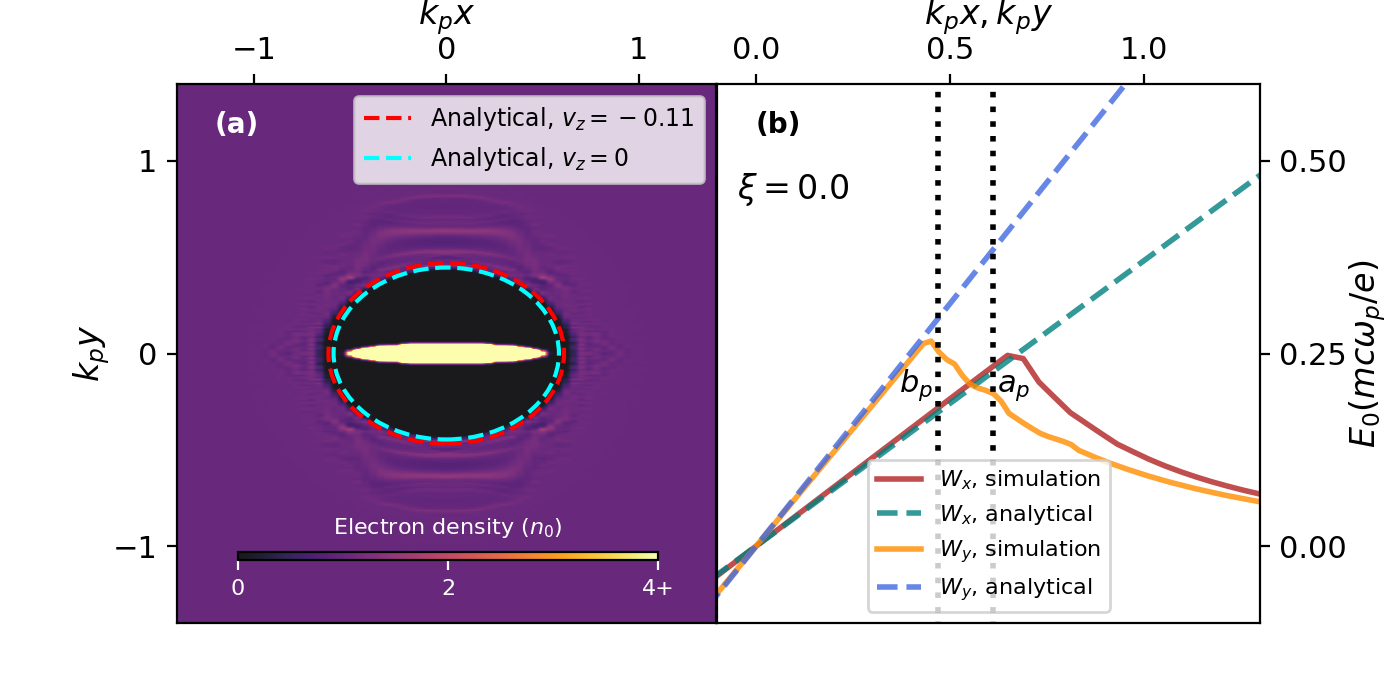}
    \caption{Long beam: Analytical calculation for the blowout shape using a beam with $n_b = 10$, $a = 0.5$, $b = 0.05$ (a) Transverse wakefield lineouts of the wake, calculated using the predicted blowout boundary (b)}
\end{figure}
\begin{figure}[H]
    \centering
    \includegraphics[width=\textwidth]{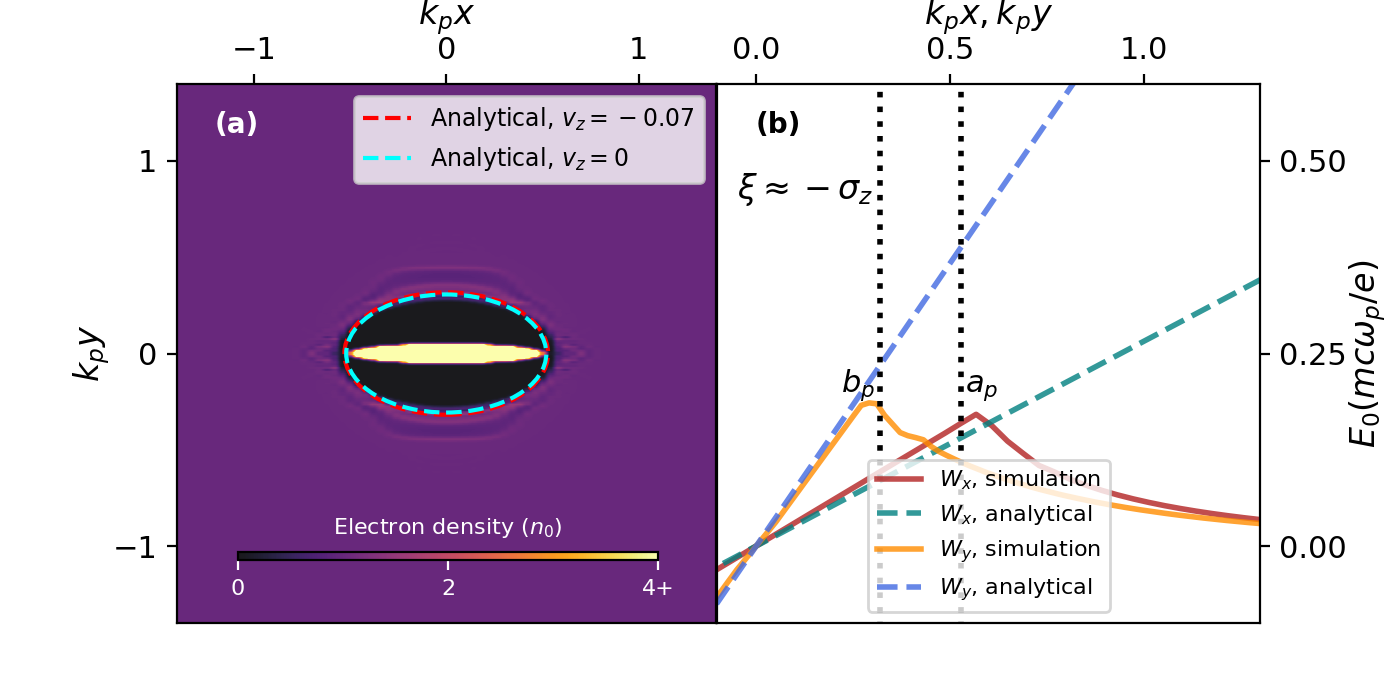}
    \caption{Long beam: Analytical calculation for the blowout shape using a beam with $n_b = 10$, $a = 0.5$ , $b = 0.05$  (a) Transverse wakefield lineouts of the wake, calculated using the predicted blowout boundary (b)}
\end{figure}
\begin{figure}[H]
    \centering
    \includegraphics[width=\textwidth]{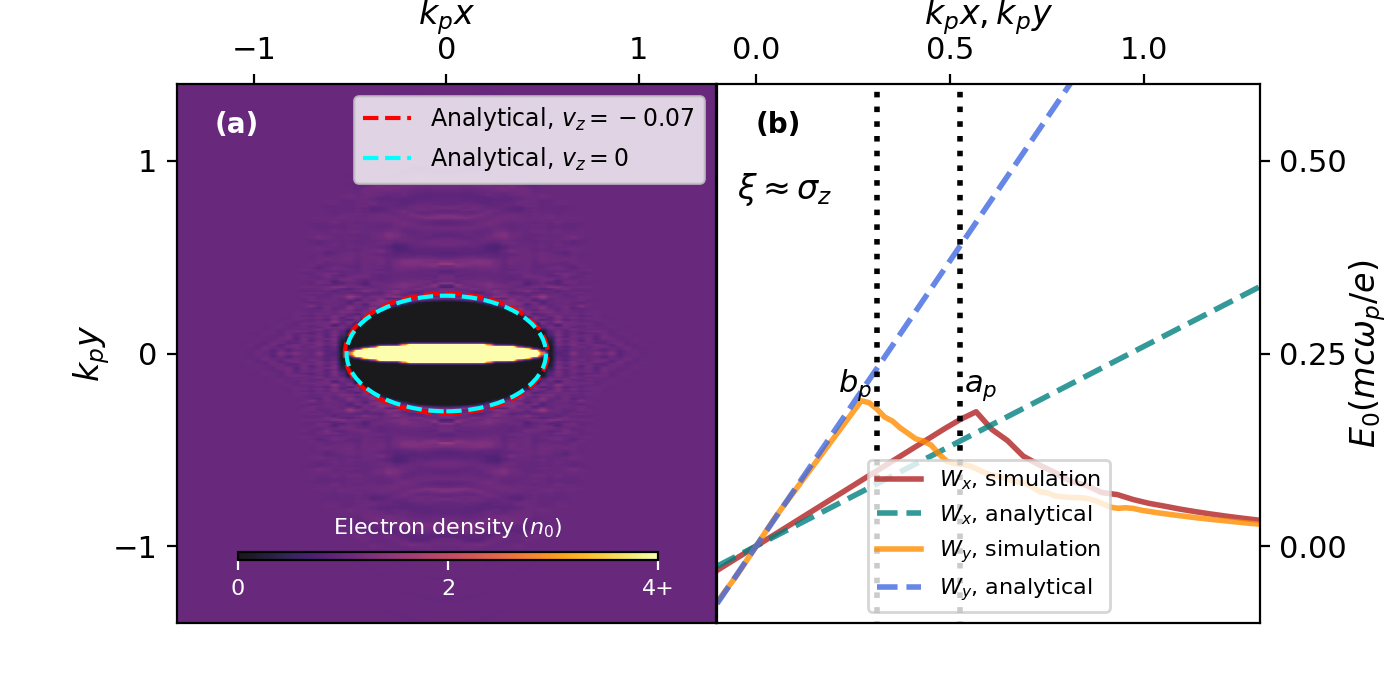}
    \caption{Long beam: Analytical calculation for the blowout shape using a beam with $n_b = 10$, $a = 0.5$ , $b = 0.05$  (a) Transverse wakefield lineouts of the wake, calculated using the predicted blowout boundary (b)}
\end{figure}

\begin{figure}[H]
    \centering
    \includegraphics[width=\textwidth]{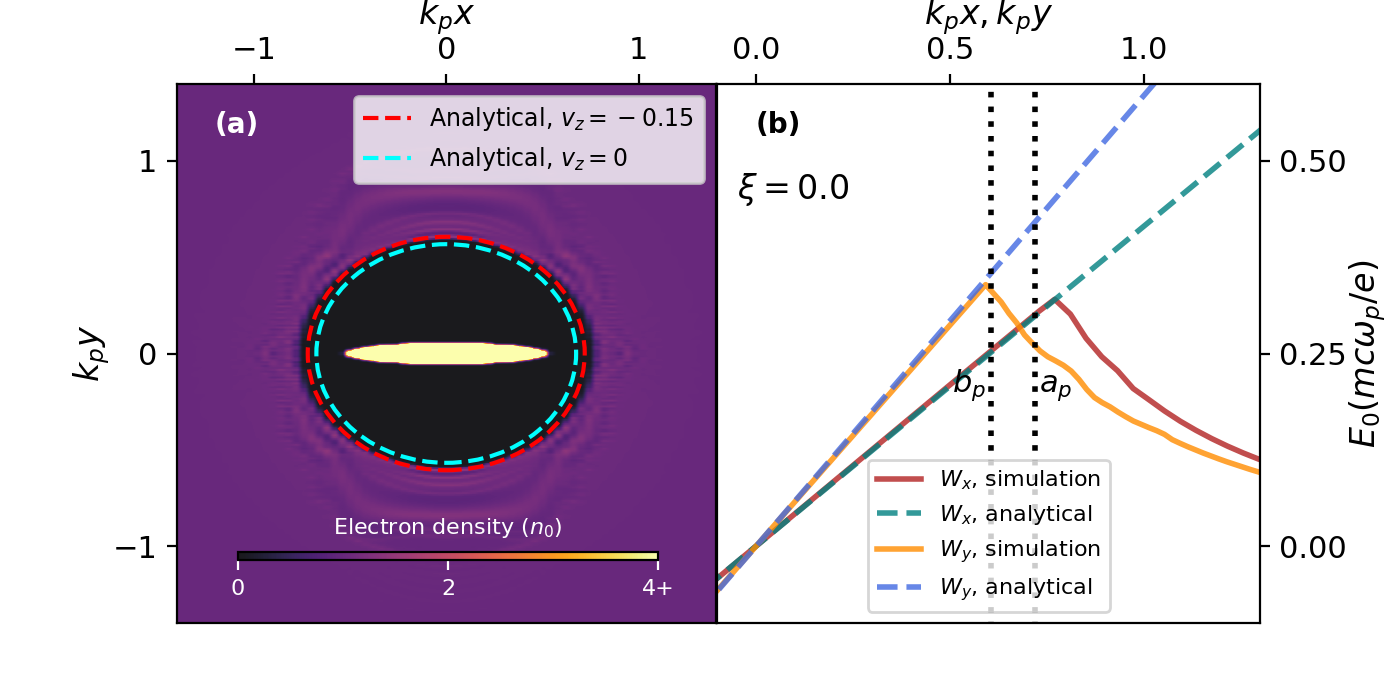}
    \caption{Long beam: Analytical calculation for the blowout shape using a beam with $n_b = 15$, $a = 0.5$ , $b = 0.05$  (a) Transverse wakefield lineouts of the wake, calculated using the predicted blowout boundary (b)}
\end{figure}
\begin{figure}[H]
    \centering
    \includegraphics[width=\textwidth]{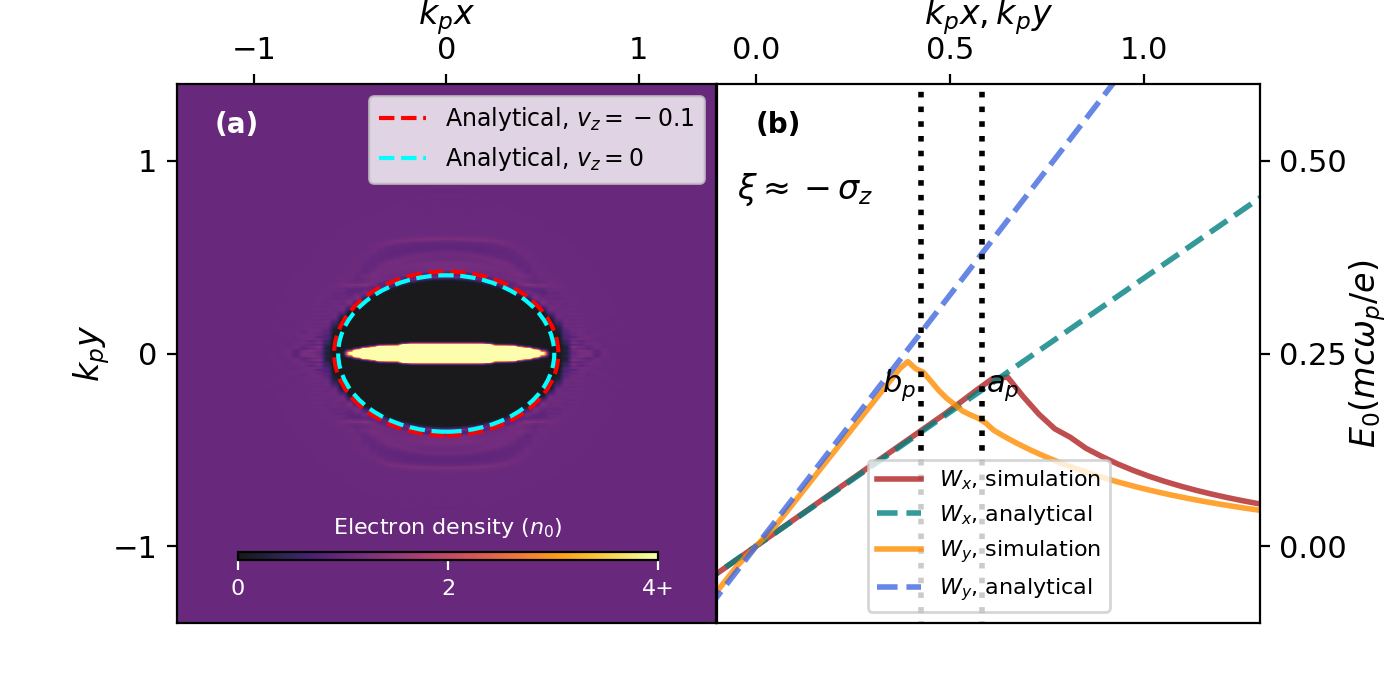}
    \caption{Long beam: Analytical calculation for the blowout shape using a beam with $n_b = 15$, $a = 0.5$ , $b = 0.05$  (a) Transverse wakefield lineouts of the wake, calculated using the predicted blowout boundary (b)}
\end{figure}
\begin{figure}[H]
    \centering
    \includegraphics[width=\textwidth]{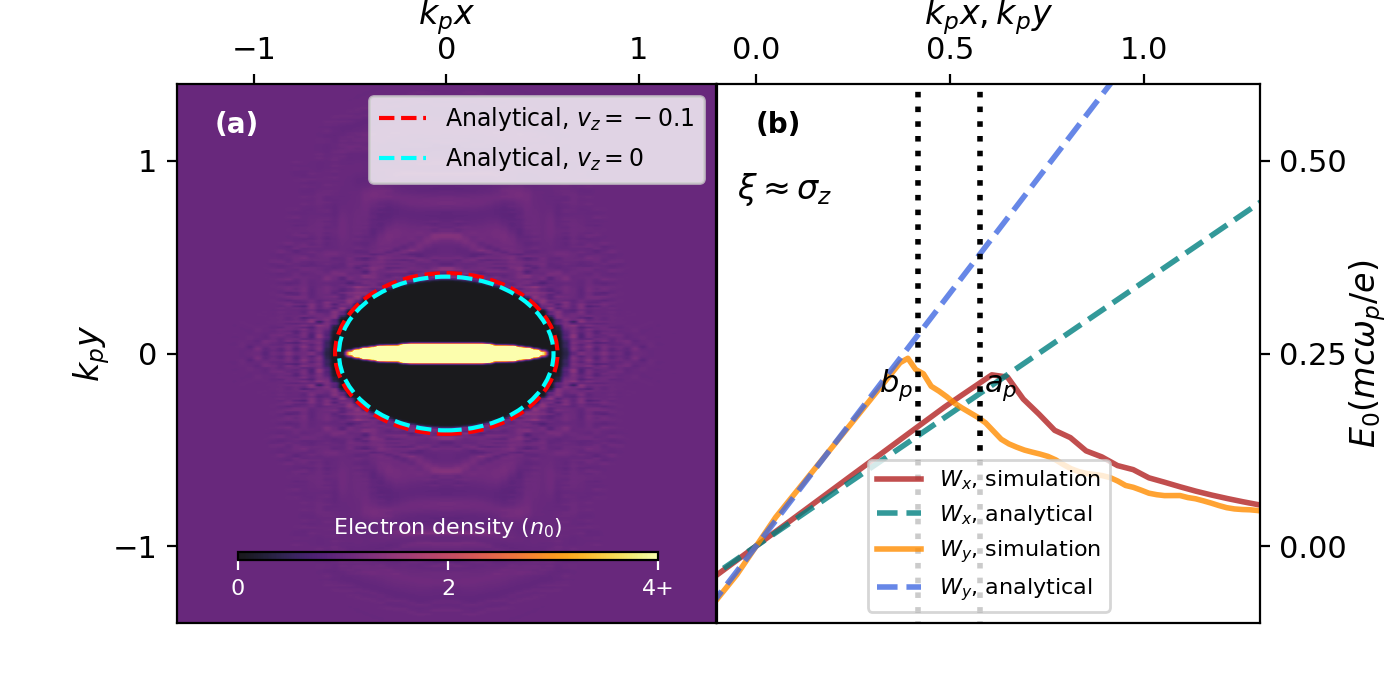}
    \caption{Long beam: Analytical calculation for the blowout shape using a beam with $n_b = 15$, $a = 0.5$ , $b = 0.05$  (a) Transverse wakefield lineouts of the wake, calculated using the predicted blowout boundary (b)}
\end{figure}

\begin{figure}[H]
    \centering
    \includegraphics[width=\textwidth]{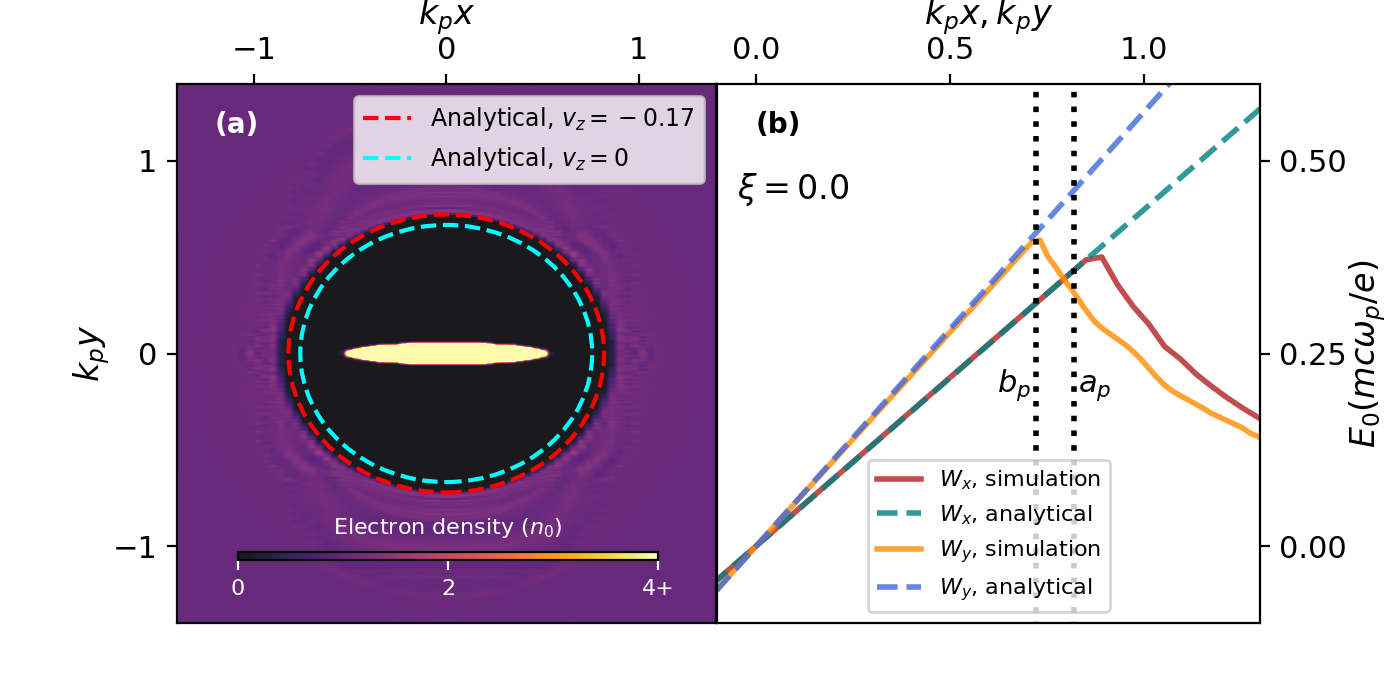}
    \caption{Long beam: Analytical calculation for the blowout shape using a beam with $n_b = 20$, $a = 0.5$ , $b = 0.05$  (a) Transverse wakefield lineouts of the wake, calculated using the predicted blowout boundary (b)}
\end{figure}
\begin{figure}[H]
    \centering
    \includegraphics[width=\textwidth]{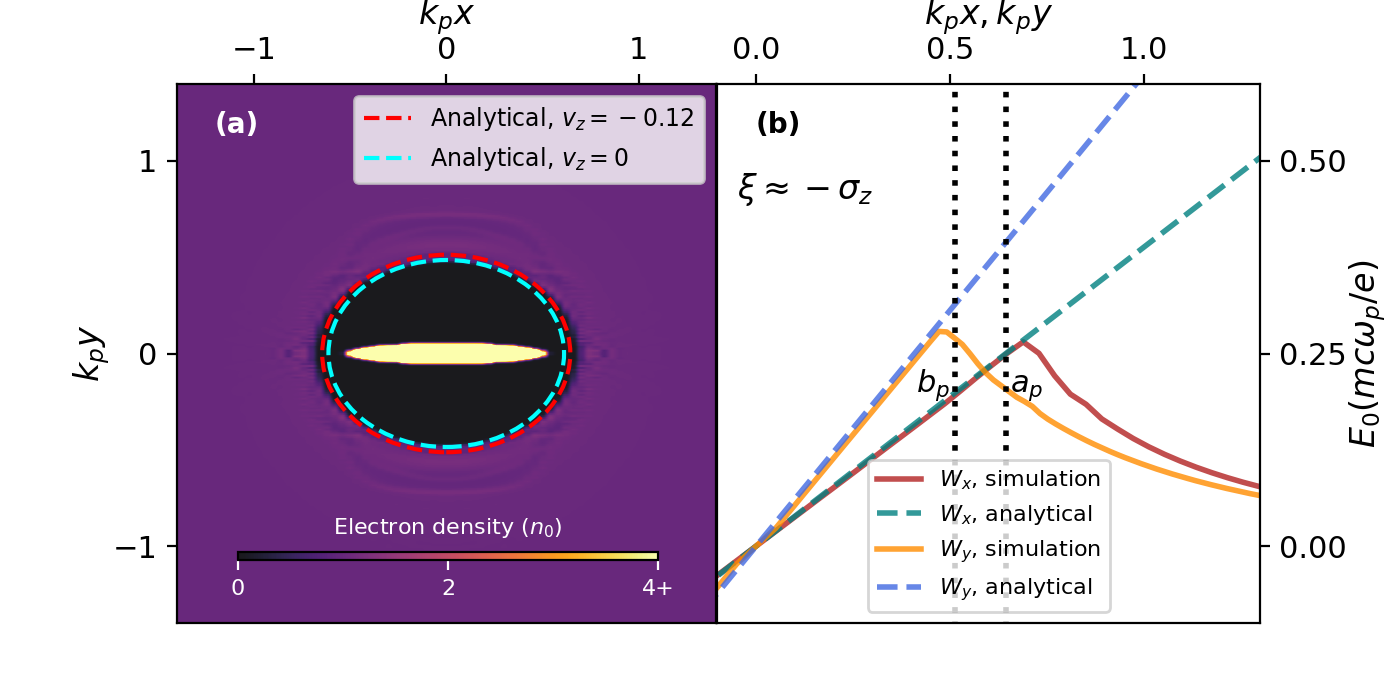}
    \caption{Long beam: Analytical calculation for the blowout shape using a beam with $n_b = 10$, $a = 0.5$ , $b = 0.05$  (a) Transverse wakefield lineouts of the wake, calculated using the predicted blowout boundary (b)}
\end{figure}
\begin{figure}[H]
    \centering
    \includegraphics[width=\textwidth]{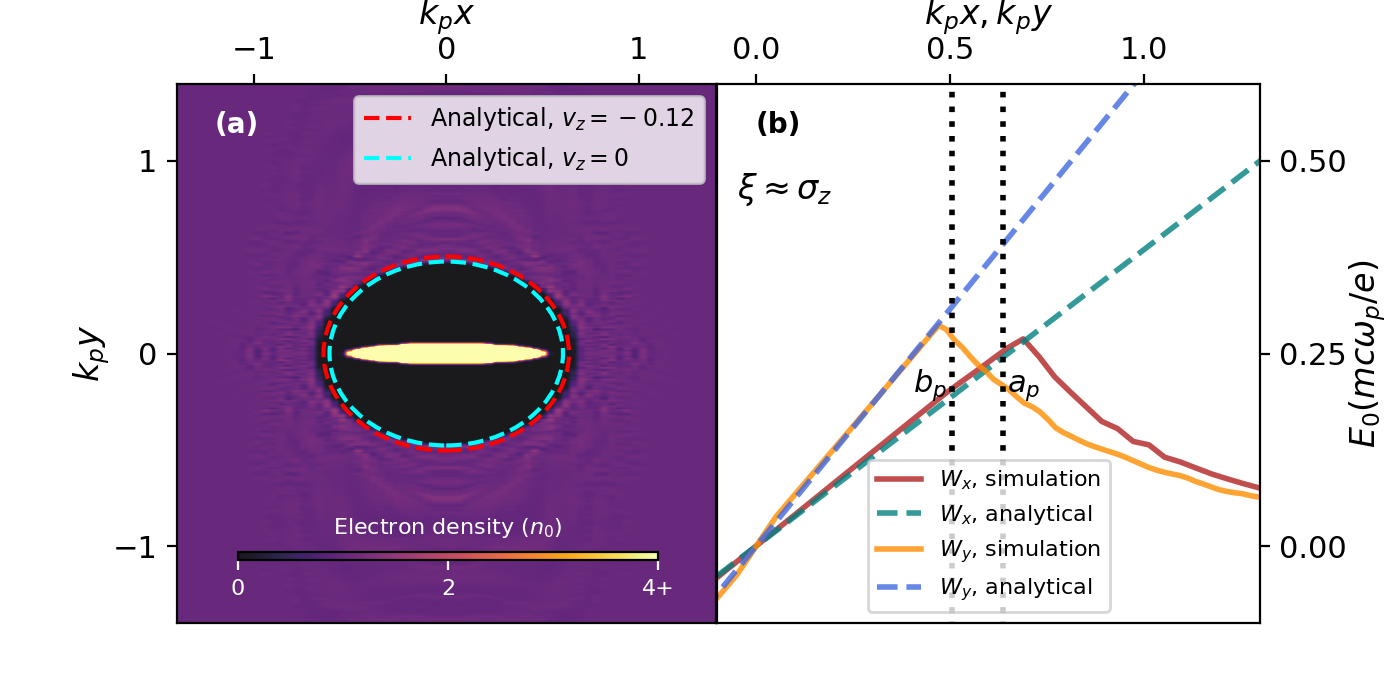}
    \caption{Long beam: Analytical calculation for the blowout shape using a beam with $n_b = 20$, $a = 0.5$ , $b = 0.05$  (a) Transverse wakefield lineouts of the wake, calculated using the predicted blowout boundary (b)}
\end{figure}

\begin{figure}[H]
    \centering
    \includegraphics[width=\textwidth]{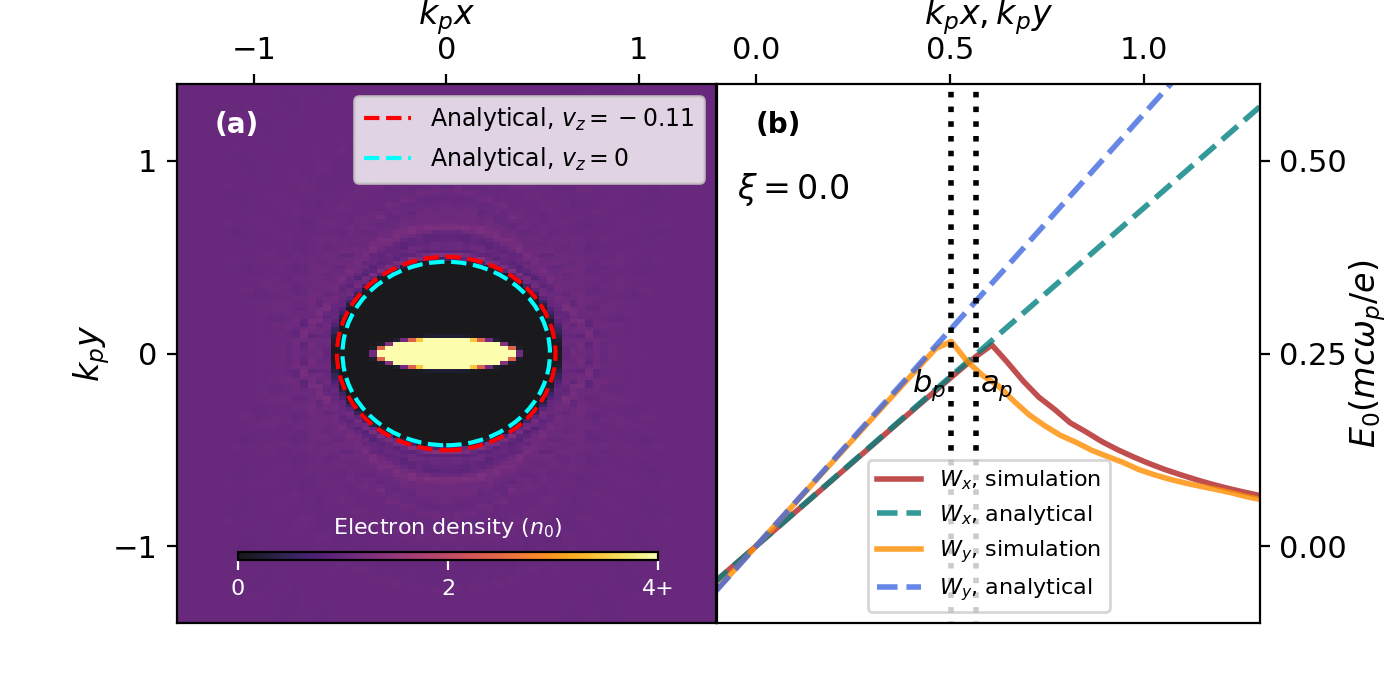}
    \caption{Long beam: Analytical calculation for the blowout shape using a beam with $n_b = 10$, $a = 0.356$ , $b = 0.0713$  (a) Transverse wakefield lineouts of the wake, calculated using the predicted blowout boundary (b)}
\end{figure}
\begin{figure}[H]
    \centering
    \includegraphics[width=\textwidth]{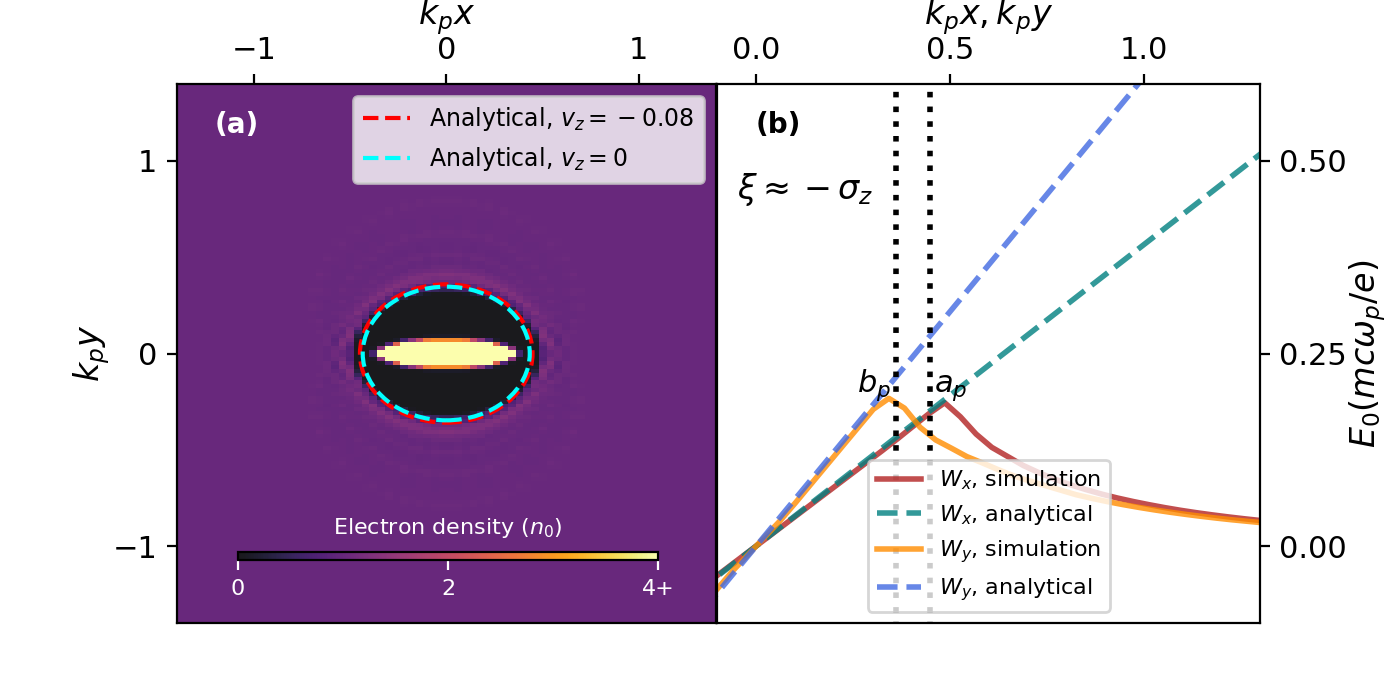}
    \caption{Long beam: Analytical calculation for the blowout shape using a beam with $n_b = 10$, $a = 0.356$ , $b = 0.0713$  (a) Transverse wakefield lineouts of the wake, calculated using the predicted blowout boundary (b)}
\end{figure}
\begin{figure}[H]
    \centering
    \includegraphics[width=\textwidth]{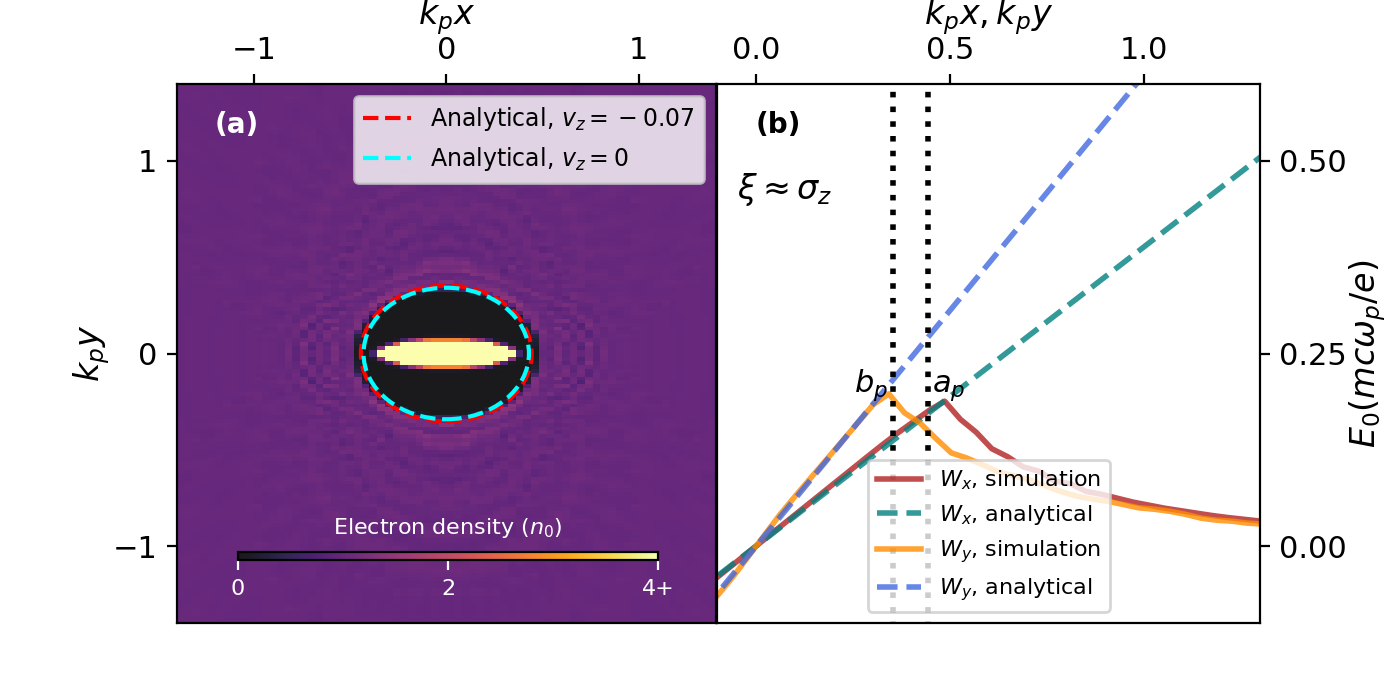}
    \caption{Long beam: Analytical calculation for the blowout shape using a beam with $n_b = 10$, $a = 0.356$ , $b = 0.0713$  (a) Transverse wakefield lineouts of the wake, calculated using the predicted blowout boundary (b)}
\end{figure}

\begin{figure}[H]
    \centering
    \includegraphics[width=\textwidth]{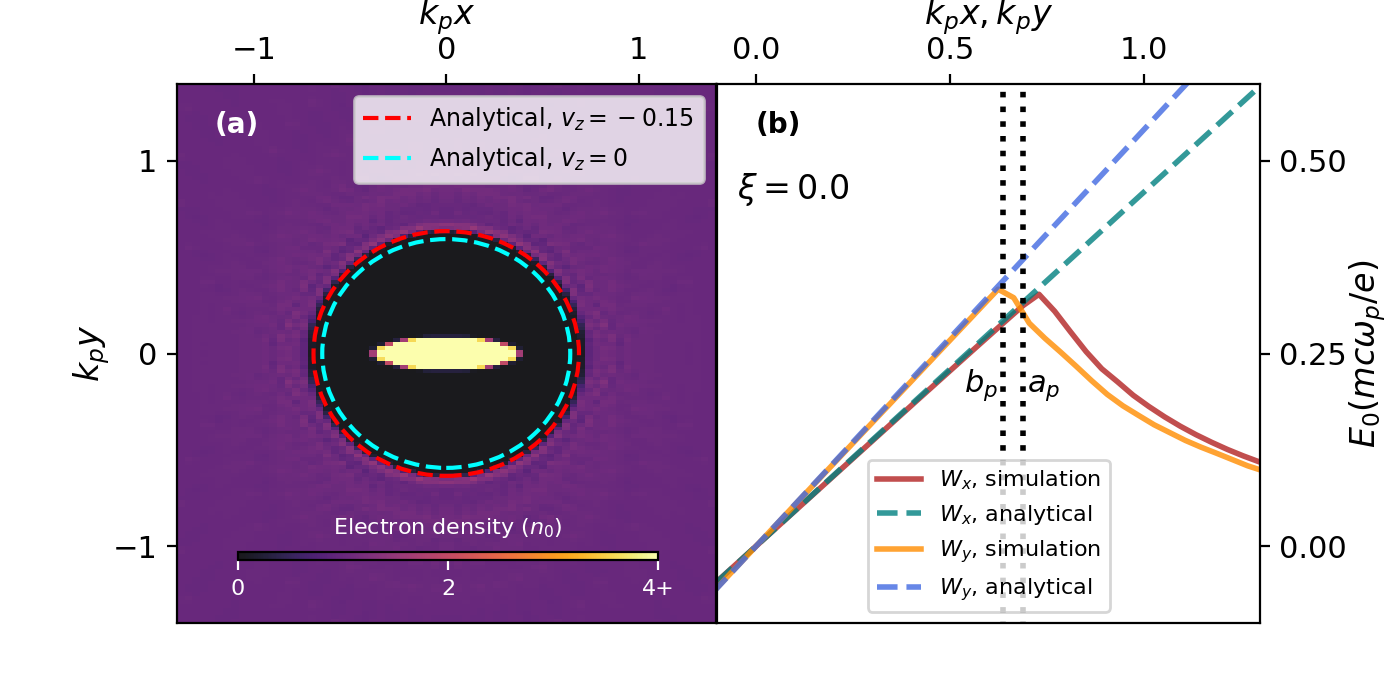}
    \caption{Long beam: Analytical calculation for the blowout shape using a beam with $n_b = 15$, $a = 0.356$ , $b = 0.0713$  (a) Transverse wakefield lineouts of the wake, calculated using the predicted blowout boundary (b)}
\end{figure}
\begin{figure}[H]
    \centering
    \includegraphics[width=\textwidth]{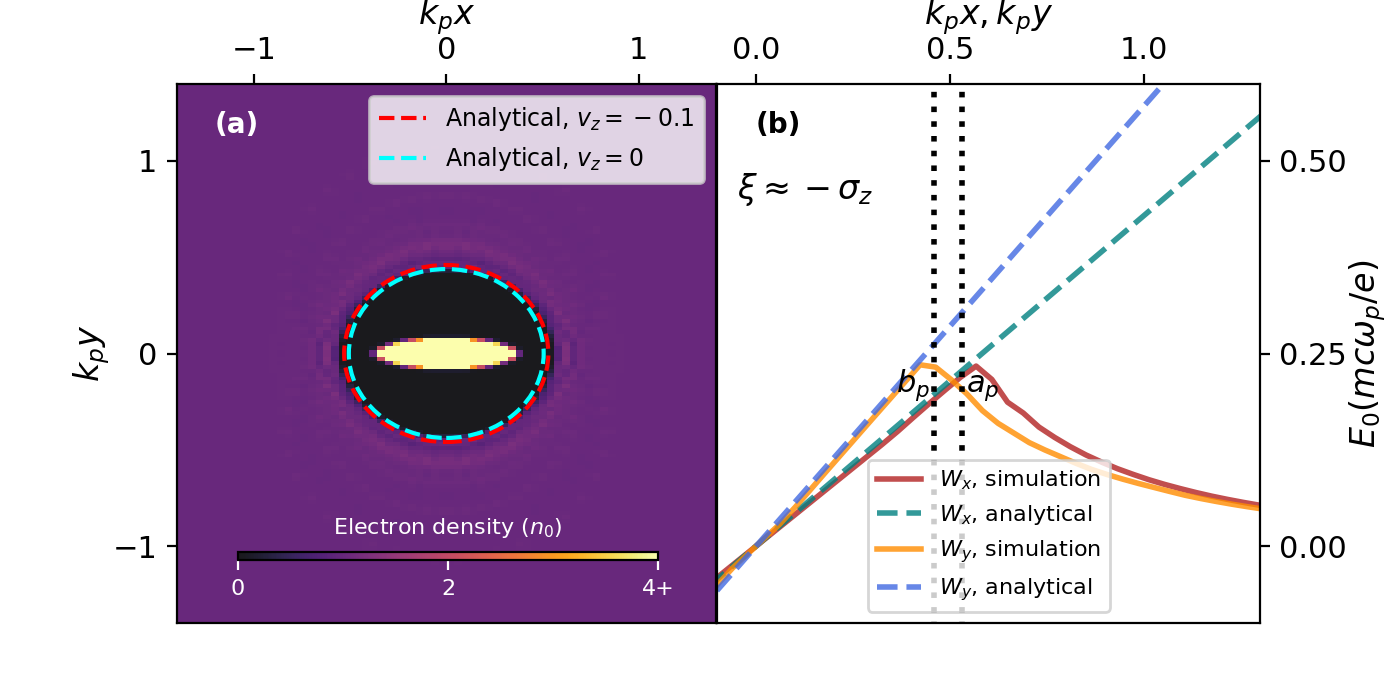}
    \caption{Long beam: Analytical calculation for the blowout shape using a beam with $n_b = 15$, $a = 0.356$ , $b = 0.0713$  (a) Transverse wakefield lineouts of the wake, calculated using the predicted blowout boundary (b)}
\end{figure}
\begin{figure}[H]
    \centering
    \includegraphics[width=\textwidth]{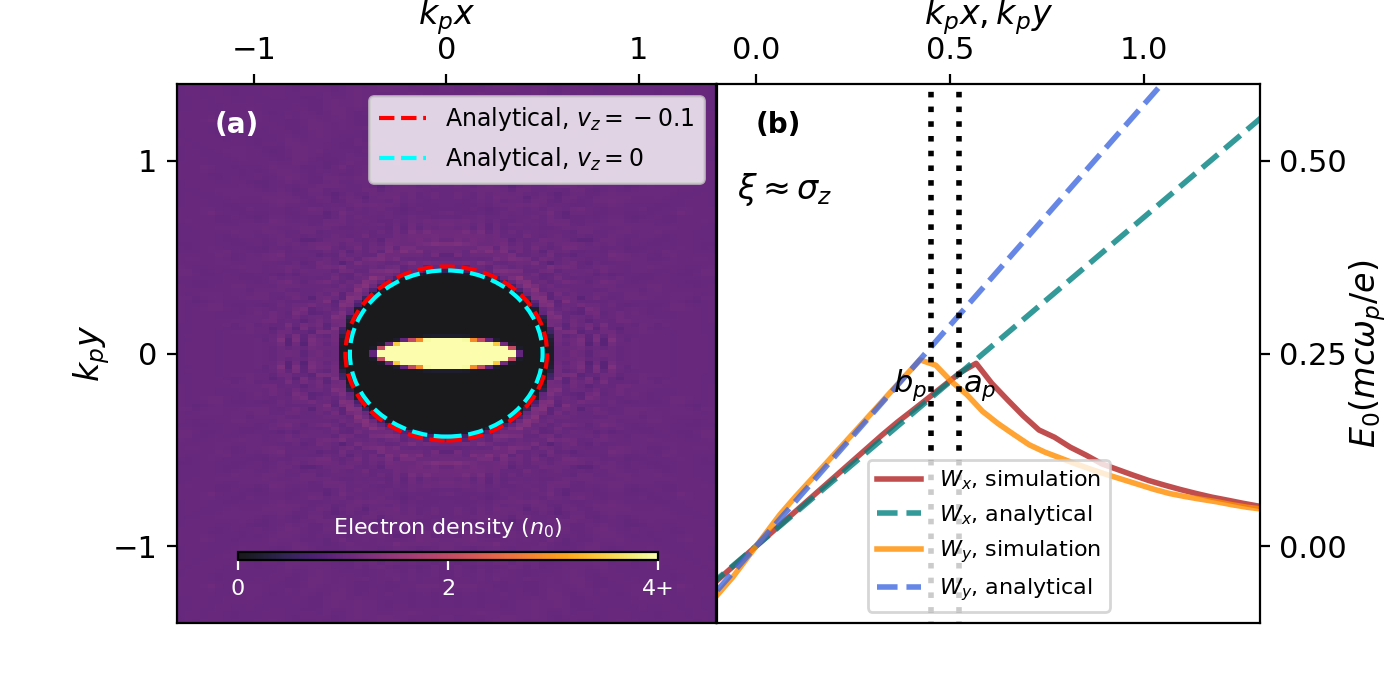}
    \caption{Long beam: Analytical calculation for the blowout shape using a beam with $n_b = 15$, $a = 0.356$ , $b = 0.0713$  (a) Transverse wakefield lineouts of the wake, calculated using the predicted blowout boundary (b)}
\end{figure}

\begin{figure}[H]
    \centering
    \includegraphics[width=\textwidth]{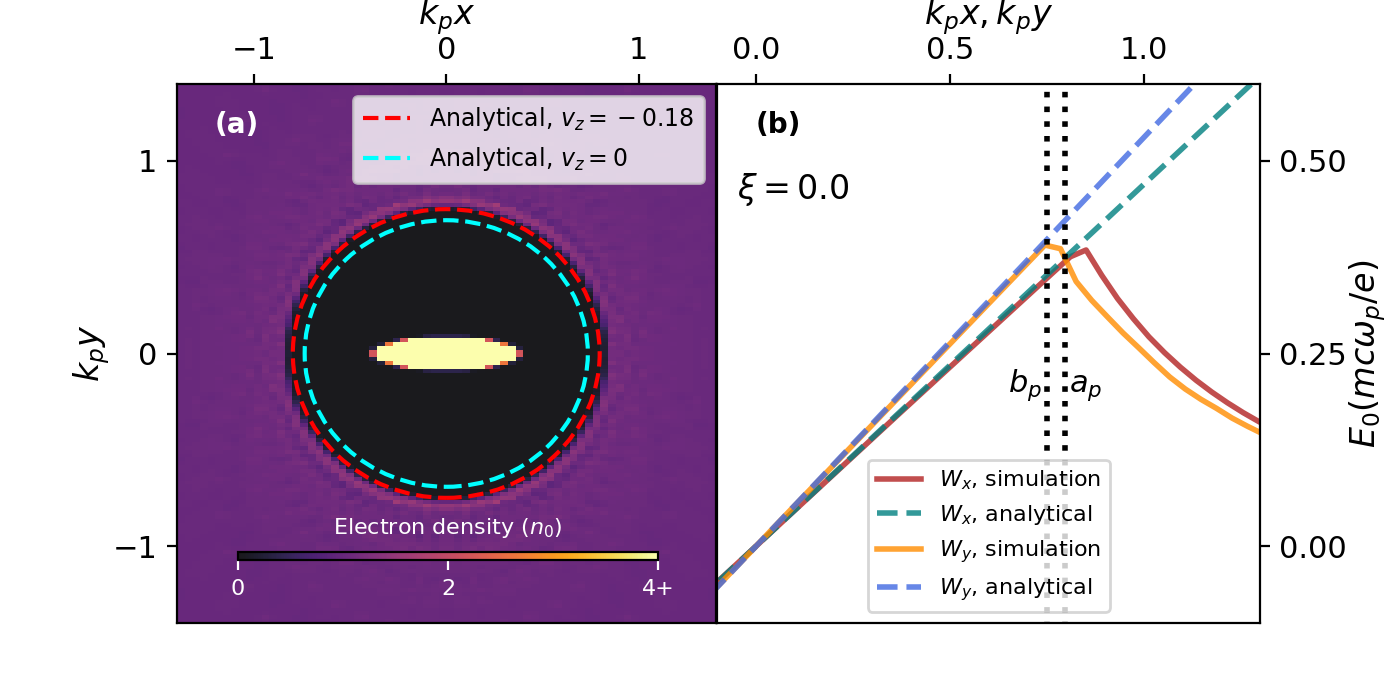}
    \caption{Long beam: Analytical calculation for the blowout shape using a beam with $n_b = 20$, $a = 0.356$ , $b = 0.0713$  (a) Transverse wakefield lineouts of the wake, calculated using the predicted blowout boundary (b)}
\end{figure}
\begin{figure}[H]
    \centering
    \includegraphics[width=\textwidth]{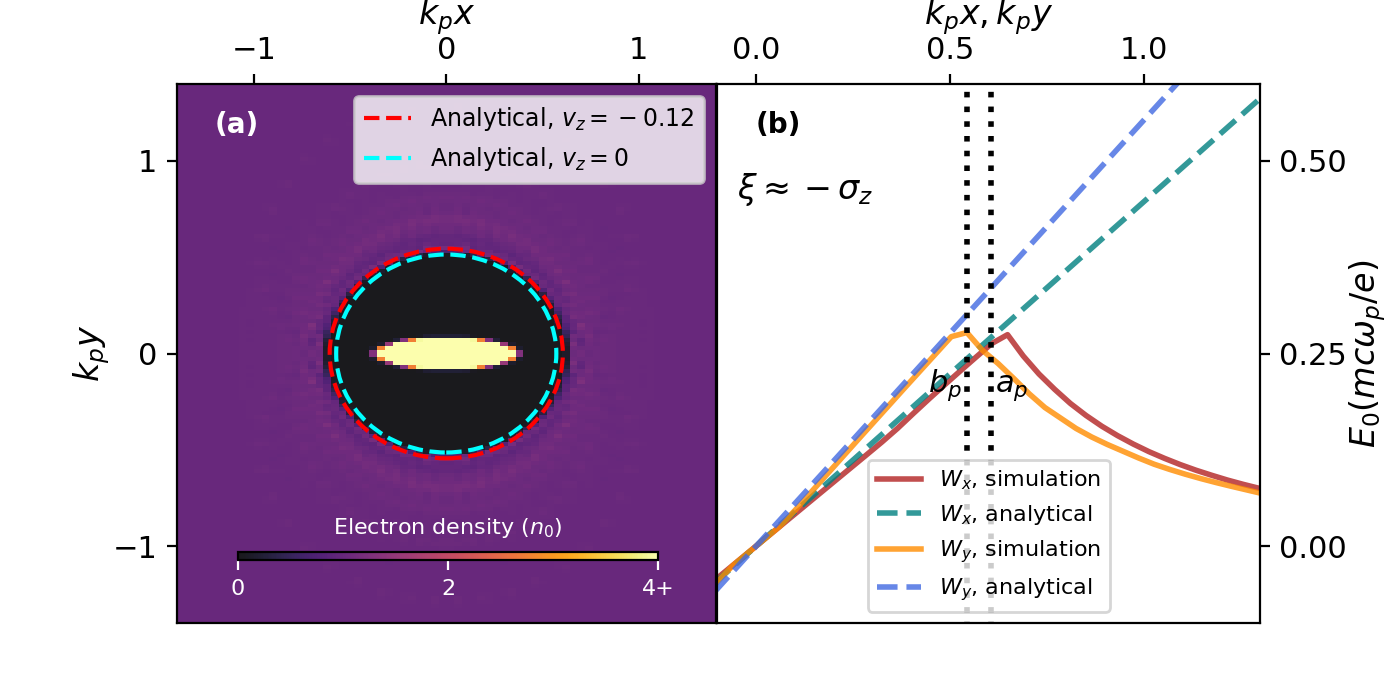}
    \caption{Long beam: Analytical calculation for the blowout shape using a beam with $n_b = 20$, $a = 0.356$ , $b = 0.0713$  (a) Transverse wakefield lineouts of the wake, calculated using the predicted blowout boundary (b)}
\end{figure}
\begin{figure}[H]
    \centering
    \includegraphics[width=\textwidth]{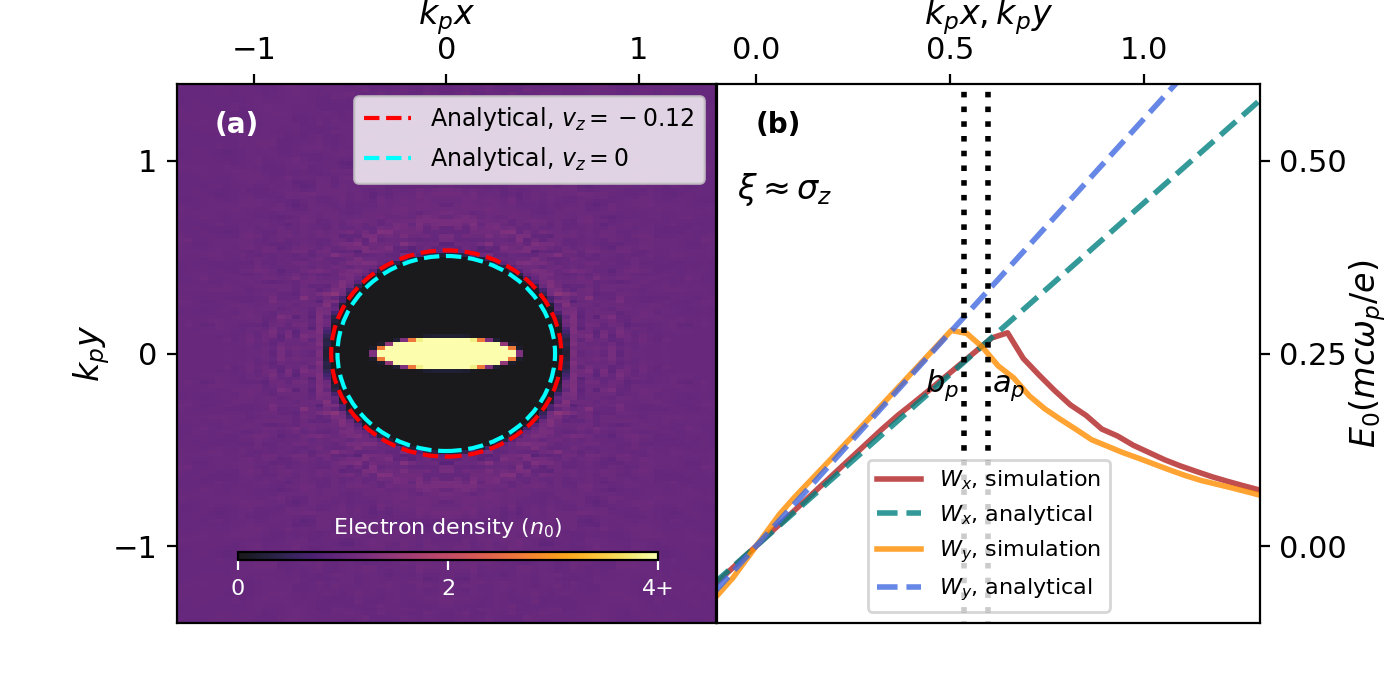}
    \caption{Long beam: Analytical calculation for the blowout shape using a beam with $n_b = 20$, $a = 0.356$ , $b = 0.0713$  (a) Transverse wakefield lineouts of the wake, calculated using the predicted blowout boundary (b)}
\end{figure}

\section{Longitudinal wakefield : Discussion and comparison to simulations}

In the Letter,  we derived the wake potential by imposing a Dirichlet boundary condition on the wake field Poisson equation, where the wake potential is set to zero at the boundary.
This approach effectively solves the same problem as considering a delta-source sheath that shields the wakefield generated by the ion column. While this method does not explicitly account for the inherent elliptical angular structure of the delta-function sheath, it captures the apparent effect of the sheath, ensuring that the electromagnetic field outside the sheath is zero. To account for the sheath, we adopted a model where the blowout boundary is discretized, and the plasma electron density outside the sheath is treated as a uniform step function with thickness, $\Delta_s$. Note that $\Delta_s$ is a free parameter that accounts for both the spike in the electron density and the exponentially decaying linear response, and is treated phenomenologically in the Letter. This simplification accurately captures the behavior of transverse fields, which depend solely on individual 2D transverse slices, parameterized by $a_p$, $b_p$ and $\Delta_s$. The model with a thin sheath ($\Delta_s \approx 0.2$) works well for transverse wakefield, while a thick sheath ($\Delta_s \approx 0.6$) works well for the longitudinal wakefield. This disparity in the sheath dependence of the wakefields likely arises from the specific shape of the exponentially decaying $\rho$ and $J_z$ distributions, and will need to be investigated in the future. 

Notably, the discrepancies between our theory and simulation become more pronounced as the blowout size diminishes or approaches the start or end of the blowout, due to the increasing significance of the actual structure of the plasma electrons and their current, which cannot be approximated as step functions any more in these cases. These longitudinally dependent errors are further amplified when derivatives of the elliptical boundaries ($a_p'$, $b_p'$ and $\Delta_s'$) are taken to evaluate longitudinal fields. where the $'$ indicates derivative with respect to $\xi$. To account for the longitudinally dependent complexity of the plasma electron sheath, previous studies have proposed more detailed models, such as incorporating linear gradients or exponentially decaying plasma electron densities. However, these approaches are challenging to implement in the flat beam regime and remains challenging and is left for future research. The simplifications used here leads to errors in estimating the magnitude of the longitudinal wakefield accurately. However, since the wakefields derive from a single potential, $\psi$, we can still begin to predict the transverse dependence of the longitudinal field as $\partial_\xi \nabla_\perp \psi= \partial_\perp \nabla_\xi \psi$. In the long beam limit, symmetry of the driven wakefield on either side of the central slice, ensures accurate predictions of the quadratic dependence of the longitudinal field. The short beam case is particularly interesting,  as the asymmetry between the beam-driven and free oscillation phases, alters the sign of the asymmetry while maintaining the quadratic dependence. This effect likely arises from the plasma electron current, which flips sign in the free oscillation phase of the short beam limit violating the constant uniform sheath assumption made in the Letter, but remains a topic of further study.

The magnitude of $W_z$ for different sheath thickness, and the quadratic dependence of $W_z(x,y)$ in the two limits, is shown below:

\begin{figure}[H]
    \centering
    \includegraphics[width=\textwidth]{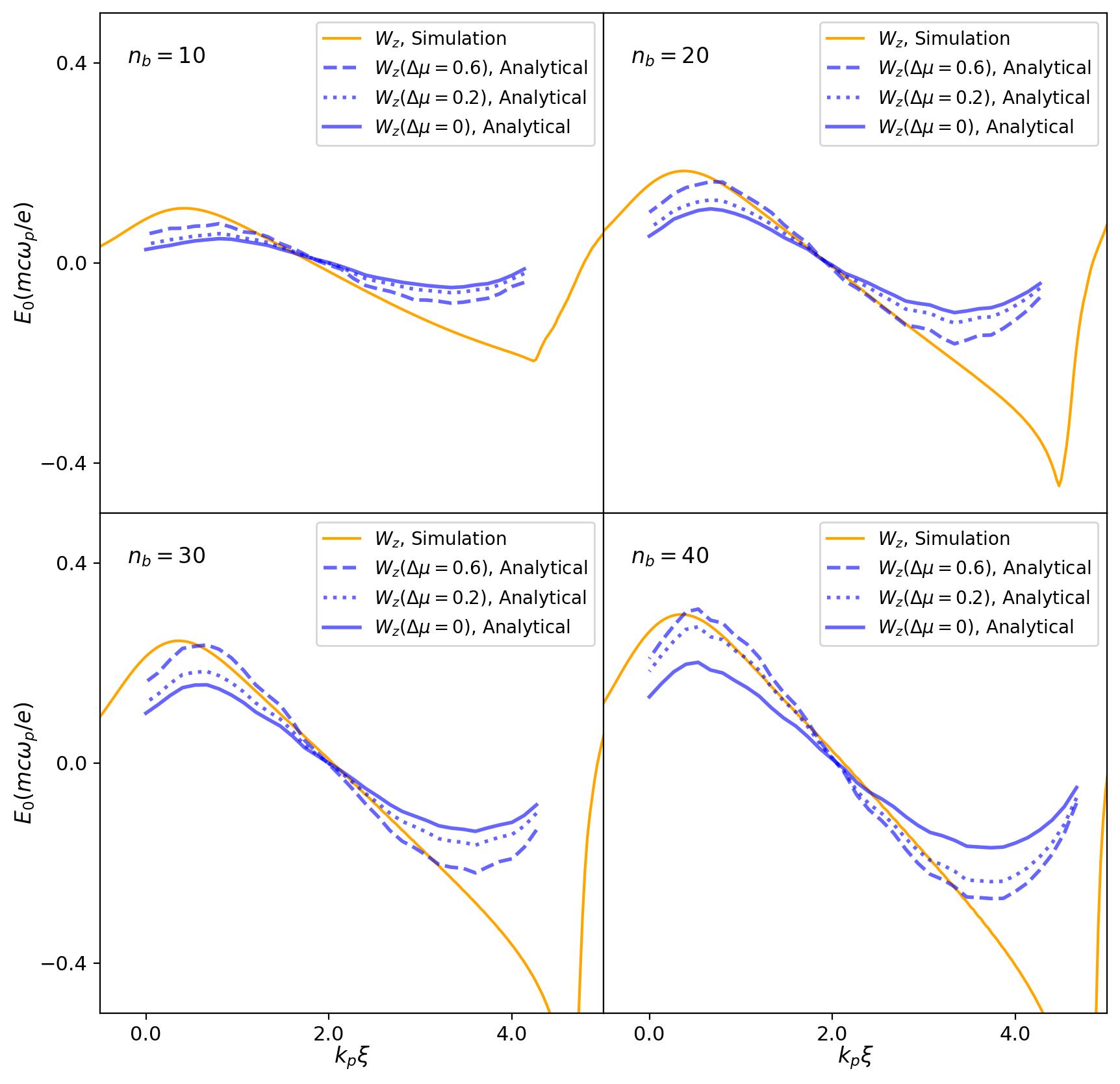}
    \caption{Short beam: Comparison between the simulation and analytical calculations for longitudinal wakefield lineouts of the wake along $\xi$, using $n_b = 10$ to $n_b = 40$, $a = 0.5$ and $b = 0.05$.}
\end{figure}

\begin{figure}[H]
    \centering
    \includegraphics[width=\textwidth]{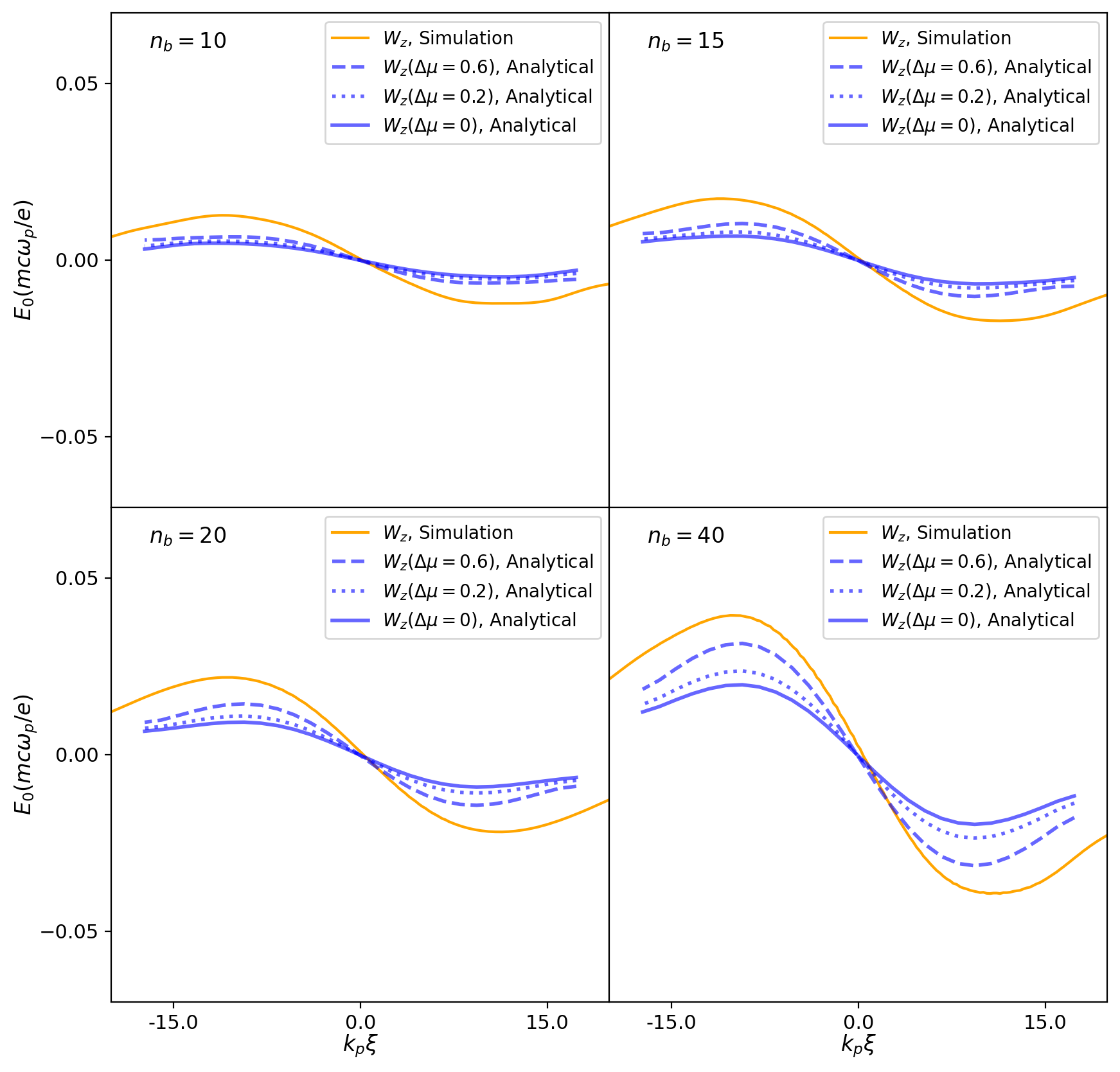}
    \caption{Long beam: Comparison between the simulation and analytical calculations for longitudinal wakefield lineouts of the wake along $\xi$, using $n_b = 10$ to $n_b = 40$, $a = 0.5$ and $b = 0.05$. Note that the sheath was added after calculating the blowout boundaries using the equation \ref{eq:equate_transverse} in the Letter.}
\end{figure}

\begin{figure}[H]
    \centering
    \includegraphics[width=\textwidth]{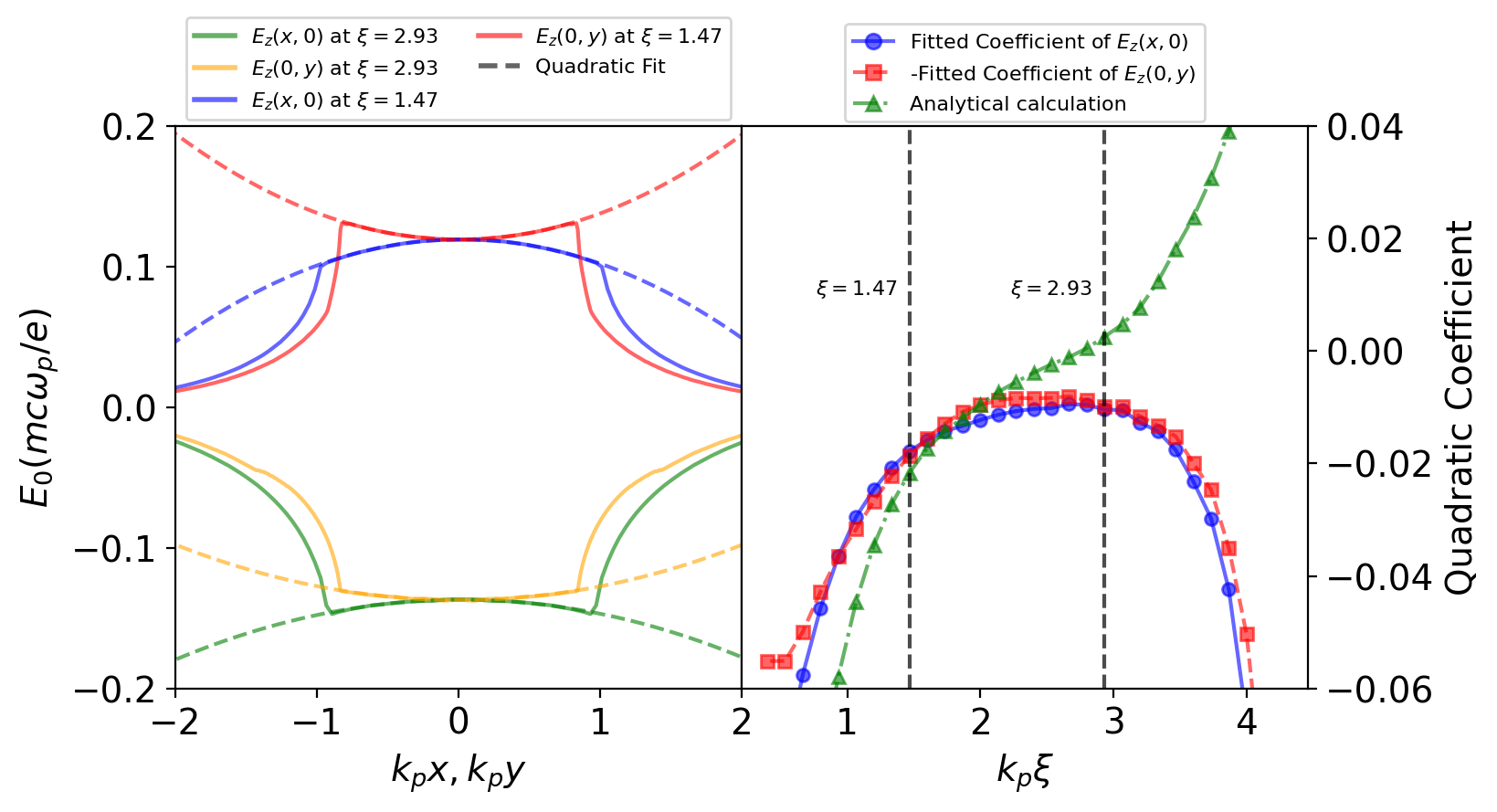}
    \caption{Short beam: Transverse lineouts of the longitudinal wakefield of the wake and quadratic fitted curve inside of the blowout (left). Fitted and predicted quadratic coefficients plotted at different $\xi$ (right).}
\end{figure}

\begin{figure}[H]
    \centering
    \includegraphics[width=\textwidth]{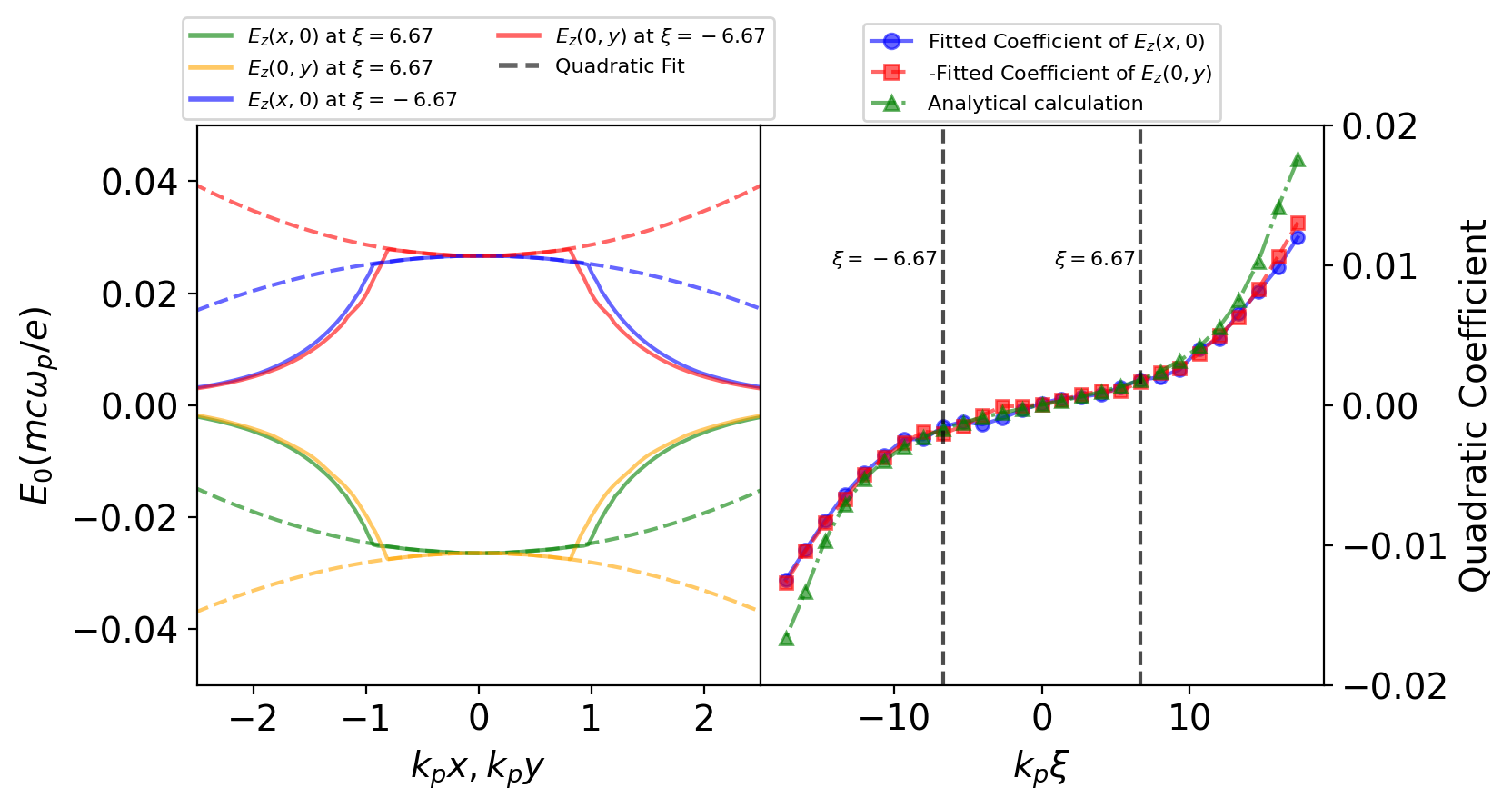}
    \caption{Long beam: Transverse lineouts of the longitudinal wakefield of the wake and quadratic fitted curve inside of the blowout (left). Fitted and predicted quadratic coefficients plotted at different $\xi$ (right).}
\end{figure}